\documentclass[prb,twocolumn,superscriptaddress,showpacs,floatfix]{revtex4-1}

\usepackage[final]{graphicx}
\usepackage{amsfonts,amssymb,amsmath}
\usepackage{prettyref}

\usepackage{braket}

\usepackage{color}
\definecolor{red}{rgb}{1.0,0.0,0.0}
\definecolor{blue}{rgb}{0.0,0.0,1.0}
\definecolor{green}{rgb}{0.0,1.0,0.0}

\newcommand{\pref}{\prettyref}

\newrefformat{cha}{Chapter~\ref{#1}}
\newrefformat{sec}{Section~\ref{#1}}
\newrefformat{fig}{Fig.~\ref{#1}}
\newrefformat{eq}{Eq.~(\ref{#1})}
\newrefformat{tab}{Tab.~\ref{#1}}
\newrefformat{app}{Appendix~\ref{#1}}

\newcommand{\etal}{\emph{et al.}}

\newcommand{\mrm}[1]{\ensuremath{\mathrm{#1}}}

\newcommand{\Tr}{\mathrm{Tr}}

\newcommand{\bmm}[1]{\begin{bmatrix}#1\end{bmatrix}}

\newcommand{\hc}[1]{\ensuremath{#1^\dagger}}

\newcommand{\ti}{\tilde}

\newcommand{\bs}[1]{\ensuremath{\boldsymbol{#1}}}
\newcommand{\cd}{\cdot}

\newcommand{\al}[1]{\begin{align} #1 \end{align}}

\newcommand{\ml}[1]{\begin{multline} #1 \end{multline}}
\newcommand{\bT}[1]{\ensuremath{\left\{ #1 \right\}}}
\newcommand{\bP}[1]{\ensuremath{\left( #1 \right)}}
\newcommand{\bS}[1]{\ensuremath{\left[ #1 \right]}}

\newcommand{\e}[1]{e^{#1}}

\newcommand{\abs}[1]{\lvert #1 \rvert}

\newcommand{\pd}[1]{\partial_{#1}}

\newcommand{\hb}{\hbar}

\newcommand{\om}{\omega}
\newcommand{\omm}[1]{\omega_{#1}}
\newcommand{\sig}{\sigma}
\newcommand{\sigg}[1]{\sigma_{#1}}

\newcommand{\asig}[1]{\braket{\sigma_{#1}(t)}}

\newcommand{\ttfct}[2]{\braket{#1(t+\tau)#2(t)}}

\begin{document}

\title{Decoherence in semiconductor cavity QED systems due to phonon couplings}

\author{P. Kaer}\email{per.kaer@gmail.com}
\affiliation{DTU Fotonik, Department of Photonics Engineering, Technical University of Denmark, Building 345, 2800 Kgs. Lyngby, Denmark}
\author{J. M{\o}rk}
\affiliation{DTU Fotonik, Department of Photonics Engineering, Technical University of Denmark, Building 345, 2800 Kgs. Lyngby, Denmark}

\date{\today}

\begin{abstract}
We investigate the effect of electron-phonon interactions on the coherence properties of single photons emitted from a  semiconductor cavity QED (Quantum Electro-Dynamics) system, i.e. a quantum dot embedded in an optical cavity. The degree of indistinguishability, governing the quantum mechanical interference between two single photons, is calculated as a function of important parameters describing the cavity QED system and the phonon reservoir, e.g. cavity quality factor, light-matter coupling strength, temperature and phonon lifetime. We show that non-Markovian effects play an important role in determining the coherence properties for typical parameter values and establish the conditions under which a Markovian approximation may be applied. The calculations are performed using a recently developed second order perturbation theory, and the limits of validity are established by comparing to an exact diagonalization approach. We find that for large cavity decay rates the perturbation theory may break down.
\end{abstract}


\maketitle




\section{Introduction}
The prospect of realizing an all-optical quantum computer \cite{Knill2001} is an important motivation for conducting research on semiconductor cavity QED (cQED) systems.
Besides single-photon detectors and standard linear optical elements, like beam-splitters and phase shifters, a main requirement for the realization of an all-optical quantum computer using the linear optics scheme\cite{Knill2001}is a single-photon source that emits near-perfect single photons on-demand and with high efficiency.
Near-perfect refers to the fact that the photons are quantum mechanically indistinguishable, in the sense that all properties of two subsequently emitted photons are exactly the same.
Under these stringent conditions, two indistinguishable single-photons will interfere quantum mechanically on a beam-splitter and display perfect bunching behavior, first observed experimentally by Hong, Ou, and Mandel (HOM) \cite{Hong1987}.
The HOM interference effect is central to the scheme proposed in \cite{Knill2001} for realizing a quantum computer and the visibility associated with the HOM interference effect is thus an important measure for the applicability of a single-photon source in a linear optical quantum computer.

Maintaining phase coherence between two subsequently emitted photons is perhaps the most difficult aspect of realizing a source of identical photons. Any process that destroys the phase coherence will thus introduce some degree of distinguishability of the emitted photons and the visibility of the HOM interference will not be 100 $\%$.
The semiconductor solid-state matrix in which the cQED system is realized, gives rich opportunities for disrupting the phase of the emitted photons through, e.g., the interactions with other carriers or quantized lattice vibrations, i.e. phonons.
In the low temperature and low excitation regime where single-photons sources are expected to operate, the main source of phase disruptions (or dephasing) stems from the interaction with phonons.\cite{Borri2001,Besombes2001}

The importance of phonons in semiconductor cQED is by now well established, both experimentally \cite{Hohenester2009c,Calic2011,Ulrich2011,Weiler2012,Madsen2013} and theoretically \cite{Wilson-Rae2002,Milde2008,Hohenester2010,Kaer2010,Majumdar2011,Kaer2012c,Glassl2012}.
However, theoretical studies of the degree of indistinguishability of photons emitted from a cQED system interacting with a non-Markovian phonon reservoir only appeared recently \cite{Kaer2013,Kaer2013a}.
Earlier theoretical studies made simplifying assumptions when considering the dynamics, such as neglecting the cavity \cite{Nazir2009b}, neglecting the non-Markovian nature of the phonons \cite{Close2012}, or not employing a microscopic model for the phonons \cite{Santori2009}. These approximations may be justified in some situations, but an extensive theoretical analysis of the system was not yet undertaken.

Experimental demonstrations of indistinguishable single photons from semiconductor QDs \cite{Santori2002,Varoutsis2005,Madsen2011} typically employ the Purcell effect to combat decoherence and usually the QD is excited through above band or quasi-resonant (p-shell) pumping.
This introduces a timing jitter in the photon emission time, which is detrimental in achieving perfect temporal overlap, which further decreases the indistinguishability.
Surprisingly, it was demonstrated that a weak pumping of the wetting layer along with p-shell pulsed excitation resulted in higher efficiency and improved indistinguishability \cite{Gazzano2013}, presumably because pumping of the wetting layer stabilizes the electrostatic environment surrounding the QD.
Recently, electrically pulsed sources of indistinguishable sources were demonstrated \cite{Patel2010}, although the requirement of post selecting photons for two-photon interference limited the overall efficiency. Pulsed resonant optical excitation of QDs \cite{He2013,Matthiesen2013,He2013b,Gao2013} led to the observation of very high degrees of single photon indistinguishability, presumably due to the complete absence of timing jitter and a very pure excitation of the QD s-shell.
However, none of the experiments employing resonant excitation made use of any significant cavity effects, resulting in an overall low efficiency.
It should be noted that while the interaction with phonons probably is the most fundamental source of decoherence in solid-state cQED systems, several of the above mentioned experimental studies quoted spectral diffusion/wandering as the major limitation in achieving indistinguishable photons under the experimental conditions considered and with the fabrication technology employed.
Spectral diffusion typically arises due to charges trapped in the vicinity of the QD, causing the QD energy levels to fluctuate on a time scale of several ns, which is much slower than fluctuations induced by the phonons.
For this reason, the highest degree of single photon interference is achieved for two subsequently emitted photons, with a temporal separation less than the time scale of slow charge-induced fluctuations.
We expect that the continuing advancement in fabrication and experimental techniques, will eventually overcome the problem of spectral diffusion and only leave fundamental decoherence processes, such as phonon scattering.

In this paper we study the influence of the non-Markovian electron-phonon interaction on the degree of indistinguishability of single photons emitted from semiconductor cQED systems.
We perform an extensive investigation of important parameters pertaining to both the phonon reservoir and the cQED system.
The present paper complements recent studies \cite{Kaer2013,Kaer2013a} on the same subject.
References \onlinecite{Kaer2013,Kaer2013a} employed a highly accurate exact diagonalization (ED) method, which treated the electron-phonon interaction on equal footing with the electron-photon interaction, thus retaining the full non-Markovian nature of the entire coupled system.
However, the ED method was numerically intensive, which effectively limited the method to low temperatures, due to the rapidly expanding phonon Hilbert space and made it difficult to explore the entire experimentally relevant parameter space.
Here we employ a recently developed non-Markovian second order theory (NM2PT) for calculating two-time correlation functions for systems in contact with non-Markovian reservoirs \cite{Goan2010,Goan2011a}.
This method provides a numerically more efficient way of analyzing the system, including the treatment of elevated temperatures.
However, the range of validity of this approximate method was not yet established.
Here we compare the NM2PT to the ED approach, and we identify experimentally relevant parameters regimes in which the NM2PT breaks down, namely the regime where the cavity decay rate is large, compared to the QD-cavity coupling constant.
Also, the parameter regimes in which the phonon reservoir effectively becomes Markovian are identified, and are found to depend strongly on whether one is interested in studying light emitted from the cavity or the QD.
Compared to earlier studies \cite{Kaer2013,Kaer2013a}, the numerically less intensive NM2PT has enabled us to thoroughly investigate important parameter dependencies, beyond those commonly encountered in cQED.
In particular the phonon lifetime is found to play an important role, as it changes the spectral properties of the phonon reservoir at small phonon frequencies.
This qualitatively changes the parameter-dependence of the indistinguishability at small QD-cavity coupling strengths, where typical experiments are conducted, and gives rise to an optimum value of the QD-cavity coupling, which is not the case if the lifetime of the phonons is neglected.

The paper is organized as follows.
In \pref{sec:theory} we briefly introduce the formalism behind the NM2PT used for calculating the two-time correlation functions in the presence of a non-Markovian reservoir, and present the model used to describe the cQED and phonon systems.
In \pref{sec:results} we present our main results.
We start by defining the degree of indistinguishability in terms of the HOM second order correlation function, using the normalized number of coincidence clicks as the main quantity.
The limits of the NM2P are then investigated and the parameter regimes of Markovian and non-Markovian behavior are discussed.
We then perform an extensive investigation of decoherence in dependence of temperature, phonon lifetime, carrier confinement, and detuning, and the emission spectra of the cQED system is considered.
In this section we also discuss useful approximations to the full non-Markovian theory.
Finally, \pref{sec:summary} summarizes our main results and conclusions.
\section{Theory}\label{sec:theory}
\subsection{Two-time system correlation functions for non-Markovian reservoirs}
It is well-known that the celebrated quantum regression theorem \cite{Carmichael1999} (QRT) does not apply for non-Markovian reservoir interactions\cite{Alonso2005,DeVega2006,DeVega2008,Goan2010,Goan2011a,Fleming2012}.
Recently Goan \etal~developed a perturbation theory for calculating two-time correlation functions of system operators \cite{Goan2010,Goan2011a} to second order in the reservoir interaction Hamiltonian, using the timeconvolution-less (TCL) approach \cite{Breuer2002}, which will be used throughout the rest of the paper.

To set the stage for presenting the theory, we assume a general Hamiltonian of the form
\al{\label{eq:uber_H}
H=H_\mrm S+H_\mrm R+H_\mrm I,
}
where $H_\mrm S$ describes the (small) system of interest, $H_\mrm R$ is the Hamiltonian of the reservoir, and finally $H_\mrm I$ is the interaction between the system and reservoir.
For simplicity we assume that all Hamiltonians are time-independent, which is typically the case for relevant cQED systems.

Following Goan \etal \cite{Goan2010,Goan2011a} the equation of motion (EOM) for the two-time correlation function $\braket{A(t_1)B(t_2)}$ for operators $A$ and $B$ belonging to the system space becomes, under the approximation that only effects to second order in $H_\mrm I$ are included and the TCL approximation is applied,
\begin{widetext}
\ml{\label{eq:NMQRT_first}
\pd{t_1}\braket{A(t_1)B(t_2)} = \frac i\hbar\braket{\bT{[H_S,A]}(t_1)B(t_2)}\\
+\frac 1{\hbar^2}\int_{t_2}^{t_1}dt' \Tr_{SR}\bP{ \bT{\ti H_I(t'-t_1)[A,H_I]}(t_1)B(t_2)\rho_T(0) + \bT{[H_I,A]\ti H_I(t'-t_1)}(t_1)B(t_2)\rho_T(0)}\\
+\frac 1{\hbar^2}\int^{t_2}_0dt' \Tr_{SR}\bP{ \bT{\ti H_I(t'-t_1)[A,H_I]}(t_1)B(t_2)\rho_T(0) + \bT{[H_I,A]}(t_1)\bT{B\ti H_I(t'-t_2)}(t_2)\rho_T(0)},
}
\end{widetext}
where the average is defined as $\langle \cdots \rangle = \mrm{Tr}_\mrm{SR}\bS{\cdots \rho_\mrm T(0)}$. Here, $\rho_\mrm T(0)=\rho_\mrm S(0)\otimes\rho_\mrm R$ is the density matrix of the entire system, which is assumed to factorize at $t=0$, and only here, into a system part $\rho_\mrm S(0)$ and a reservoir part $\rho_\mrm R$.
The interaction picture representation of an operator $X$ is defined as
\al{\label{eq:int_picture_def}
\ti{X}(t)=U^\dag(t)XU(t), \quad U(t)=\e{-\textrm{i}(H_S + H_R)t/\hbar}.
}
We can bring the general expression \pref{eq:NMQRT_first} into a more practical form by making a few assumptions on the properties of the reservoir.
First, we assume the following form of the interaction Hamiltonian
\al{
H_{I}=\sum_{\nu_1\nu_2}P_{\nu_1\nu_2}R_{\nu_1\nu_2},
}
where $P_{\nu_1\nu_2}$ is a general system operator and $R_{\nu_1\nu_2}$ is a general reservoir operator.
We note that most physically relevant interactions can be brought into this form.
Second, we assume that the reservoir is in a stationary state such that its correlation functions only depend on difference times
\al{\label{eq:uber_D_def}
D_{\nu_1\nu_2\nu_3\nu_4}(t-t')=\mathrm{Tr}_{R}\bS{\ti R_{\nu_1\nu_2}(t-t')R_{\nu_3\nu_4}\rho_{R}}
,
}
which is the case for, e.g., a thermal state.
Third, if one assumes $\hc R_{\nu_{1}\nu_{2}}=R_{\nu_{2}\nu_{1}}$, we have the following symmetry relation
\al{
D^*_{\nu_1\nu_2\nu_3\nu_4}(t)=D_{\nu_2\nu_1\nu_4\nu_3}(-t).
}
Finally, we assume $\hc P_{\nu_{1}\nu_{2}}=P_{\nu_{2}\nu_{1}}$.
We note that these assumptions usually are fulfilled for relevant physical interactions, but are not essential and only serve to simplify the resulting equations.

Using these assumptions, employing the Born approximation for factorizing the system and reservoir degrees of freedom, and moving to a $(t+\tau,t)$-frame we get the EOM for the NM2PT
\begin{widetext}
\ml{\label{eq:super_uber_twotime_EOM}
\pd{\tau}\braket{A(t+\tau)B(t)} = \frac i\hbar\braket{\bT{[H_S,A]}(t+\tau)B(t)}\\
+\sum_{\nu_{1}\nu_{2}\nu_{3}\nu_{4}}\frac 1{\hbar^2}\int_{0}^{\tau}dt' \bS{ D^{*}_{\nu_{4}\nu_{3}\nu_{2}\nu_1}(t')  \braket{\{\ti P_{\nu_{1}\nu_{2}}(-t')[A,P_{\nu_{3}\nu_{4}}]\}(t+\tau)B(t)} \right.\\
+ \left. D_{\nu_{4}\nu_{3}\nu_{2}\nu_{1}}(t')\braket{\{[\hc P_{\nu_{3}\nu_{4}},A]\ti P^{\dagger}_{\nu_{1}\nu_{2}}(-t')\}(t+\tau)B(t)} }\\
+\sum_{\nu_{1}\nu_{2}\nu_{3}\nu_{4}}\frac 1{\hbar^2}\int^{t+\tau}_\tau dt' \bS{ D^{*}_{\nu_{4}\nu_{3}\nu_{2}\nu_1}(t') \braket{\{\ti P_{\nu_{1}\nu_{2}}(-t')[A,P_{\nu_{3}\nu_{4}}]\}(t+\tau)B(t)} \right. \\
\left. + D_{\nu_{4}\nu_{3}\nu_{2}\nu_{1}}(t')\braket{\{[\hc P_{\nu_{3}\nu_{4}},A]\}(t+\tau)\{B\ti P^{\dagger}_{\nu_{1}\nu_{2}}(-[t'-\tau])\}(t)} }.
}
\end{widetext}
Before proceeding we discuss the general features of non-Markovian effects, as they are represented in \pref{eq:super_uber_twotime_EOM}, and in which limits Markovian behavior may be expected.
The timescale on which $D_{\nu_{4}\nu_{3}\nu_{2}\nu_{1}}(t)$ decays, denoted $\tau_\mrm{corr}$ and often referred to as the correlation time or memory depth of the reservoir, in comparison to the characteristic time constants of the small system, with which the reservoir interacts, is very important for the interaction dynamics.
The system plus reservoir can typically be characterized by three timescales: The first is the coherent timescale, which is determined by the inverse difference between two typical eigenenergies of the system Hamiltonian $H_\mrm S$, thus $\tau_\mrm{coh}\sim \abs{\om_i-\om_j}^{-1}$.
The second is the overall relaxation time of the system, $\tau_\mrm{relax}$, which is determined by all the relaxation processes affecting the system.
The third timescale is the correlation or memory time of the reservoir and is set by the decay of the reservoir correlation function, which depends on the properties of the bare reservoir, such as dispersion and temperature, as well as the nature of the system-reservoir interaction Hamiltonian, $H_\mrm I$.

\begin{figure}[ht]
 \includegraphics[width=0.48\textwidth]{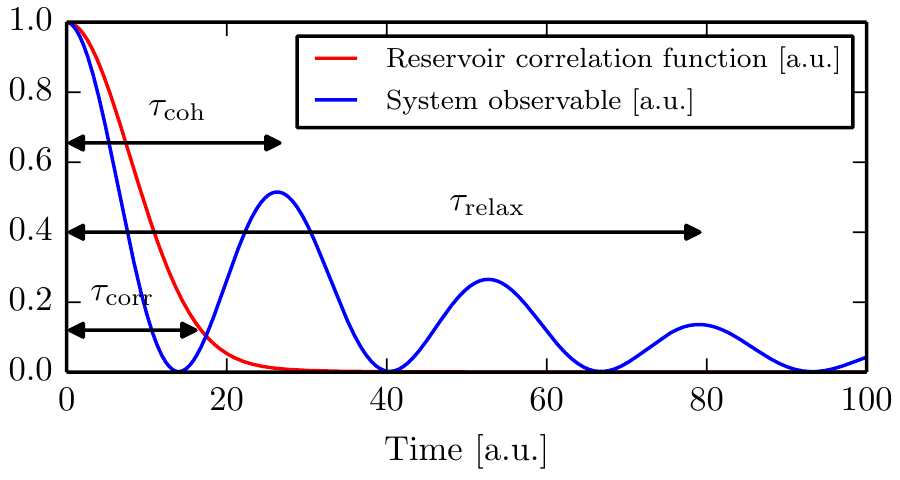}
 \caption{Illustration of the typical timescales characterizing the system and reservoir. The typical coherent and relaxation timescales of the cQED system are denoted $\tau_\mrm{coh}$ and $\tau_\mrm{relax}$, respectively. The memory time of the reservoir is denoted $\tau_\mrm{corr}$.}
\label{fig:simple_decay}
\end{figure}

All three timescales are illustrated in \pref{fig:simple_decay}.
For the kind of cQED systems we are considering, $\tau_\mrm{coh}$ would typically be determined by the QD-cavity coupling $g$ and the oscillations would be Rabi flops and $\tau_\mrm{relax}$ could, e.g., be the Purcell rate, scaling as $g^2/\kappa$, where the cavity decay rate $\kappa$ enters.

Depending on the relative values of the three timescales, three important regimes of operation for the total system can be identified.

(I) $\tau_\mrm{corr}\ll\tau_\mrm{coh},\tau_\mrm{relax}$, here reservoir correlations decay much faster than the system can react and the reservoir correlation function can be assumed to be a delta function, $D_{\nu_{4}\nu_{3}\nu_{2}\nu_1}(t') = \bar D_{\nu_{4}\nu_{3}\nu_{2}\nu_1} \delta(t')$ with $\bar D_{\nu_{4}\nu_{3}\nu_{2}\nu_1}$ being a constant.
Using this in \pref{eq:super_uber_twotime_EOM} yields
\begin{widetext}
\ml{\label{eq:QRT_EOM}
\pd{\tau}\braket{A(t+\tau)B(t)} = \frac i\hbar\braket{\bT{[H_S,A]}(t+\tau)B(t)}\\
+\sum_{\nu_{1}\nu_{2}\nu_{3}\nu_{4}}\frac 1{2\hbar^2}\bS{ \bar D^{*}_{\nu_{4}\nu_{3}\nu_{2}\nu_1}  \braket{\{P_{\nu_{1}\nu_{2}}[A,P_{\nu_{3}\nu_{4}}]\}(t+\tau)B(t)} \right.\\
+ \left. \bar D_{\nu_{4}\nu_{3}\nu_{2}\nu_{1}}\braket{\{[\hc P_{\nu_{3}\nu_{4}},A] P^{\dagger}_{\nu_{1}\nu_{2}}\}(t+\tau)B(t)} },
}
\end{widetext}
Here the scattering rates are constants, that do not depend on the properties of the small system governed by the Hamiltonian $H_\mrm S$.
In the spectral domain, a delta-correlated correlation functions translates into a constant, unstructured, spectrum, and thus transitions occurring inside the small system always experience the same reservoir density, leading to constant scattering rates.
The EOM \pref{eq:QRT_EOM} is identical to that derived within the QRT procedure and we denote this regime the \textit{Markovian regime for an unstructured reservoir}.

(II) $\tau_\mrm{corr}\approx\tau_\mrm{coh},\tau_\mrm{relax}$ and $\tau\gg\tau_\mrm{corr}$, here the reservoir correlation time is comparable to at least one of the system timescales and we consider difference times $\tau$ that are considerably larger than the reservoir correlation time.
For times larger than $\tau_\mrm{coh}$ all integration limits can be taken to infinity and the terms in the two last lines of \pref{eq:super_uber_twotime_EOM} vanish.
The long-time limit signifies that we are in the Markovian regime, but due to the assumption that at least one of the system timescales was comparable to the reservoir correlation time, the system still samples the structure of the reservoir.
Indeed, the spectrum of the reservoir now shows significant variations over the bandwidth of the system, and different transitions occurring in the system sample different reservoir densities.
This has the practical consequence that the reservoir induced scattering rates depend on the properties of the system, more specifically via the parameters of $H_\mrm S$ through the time-evolution operator $U(t)$ defining the interaction picture, see \pref{eq:int_picture_def}.
Notice though, that the rates themselves are independent of time.
The long-time limit also implies that scattering events must conserve energy, and thus the scattering rates only describe transitions between real states of the system.
We denote this regime the \textit{Markovian regime for a structured reservoir}.

(III) $\tau_\mrm{corr}\approx\tau_\mrm{coh},\tau_\mrm{relax}$ and $\tau<\tau_\mrm{corr}$, here the system and reservoir timescales are related as in regime II, but we assume that the difference time $\tau$ is smaller than the reservoir correlation time.
In this regime all terms of \pref{eq:super_uber_twotime_EOM} are potentially important and the scattering rates generally depend on time.
As we are in the short-time limit, scattering events occurring between the system and reservoir need not conserve energy and virtual transitions in general dominate the dynamics.
Thus, the entire spectrum of the reservoir is sampled and not just specific energies.
This regime is referred to as \textit{non-Markovian}.

For further discussions and examples of these regimes see \pref{sec:puredephrates} and \pref{sec:useful_approximations}.

Other relaxation processes, where the reservoir can be assumed to be unstructured, such as cavity decay, are included via terms of the Lindblad type
\begin{widetext}
\al{\label{eq:twotime_Lindblad}
\pd{\tau}\braket{A(t+\tau)B(t)}\vert_\mrm{Lindblad} = -\frac \alpha 2 \bS{ \braket{\{\hc P P A\}(t+\tau)B(t)}+\braket{ \{A\hc P P\}(t+\tau)B(t)} -2\braket{\{\hc P A P\}(t+\tau)B(t)}},
}
\end{widetext}
where $\alpha$ is the decay rate and the operator $P$ describes the corresponding transition.

As initial conditions for the two-time function $\braket{A(t+\tau)B(t)}$, the corresponding one-time function $\braket{A(t)B(t)}$ is needed.
The EOM for the one-time function is easily obtained from the general two-time EOM by setting $B=I$ and $t=0$.
\subsection{Jaynes-Cummings model with longitudinal acoustical phonons}
\begin{figure}[ht]
 \includegraphics[width=0.48\textwidth]{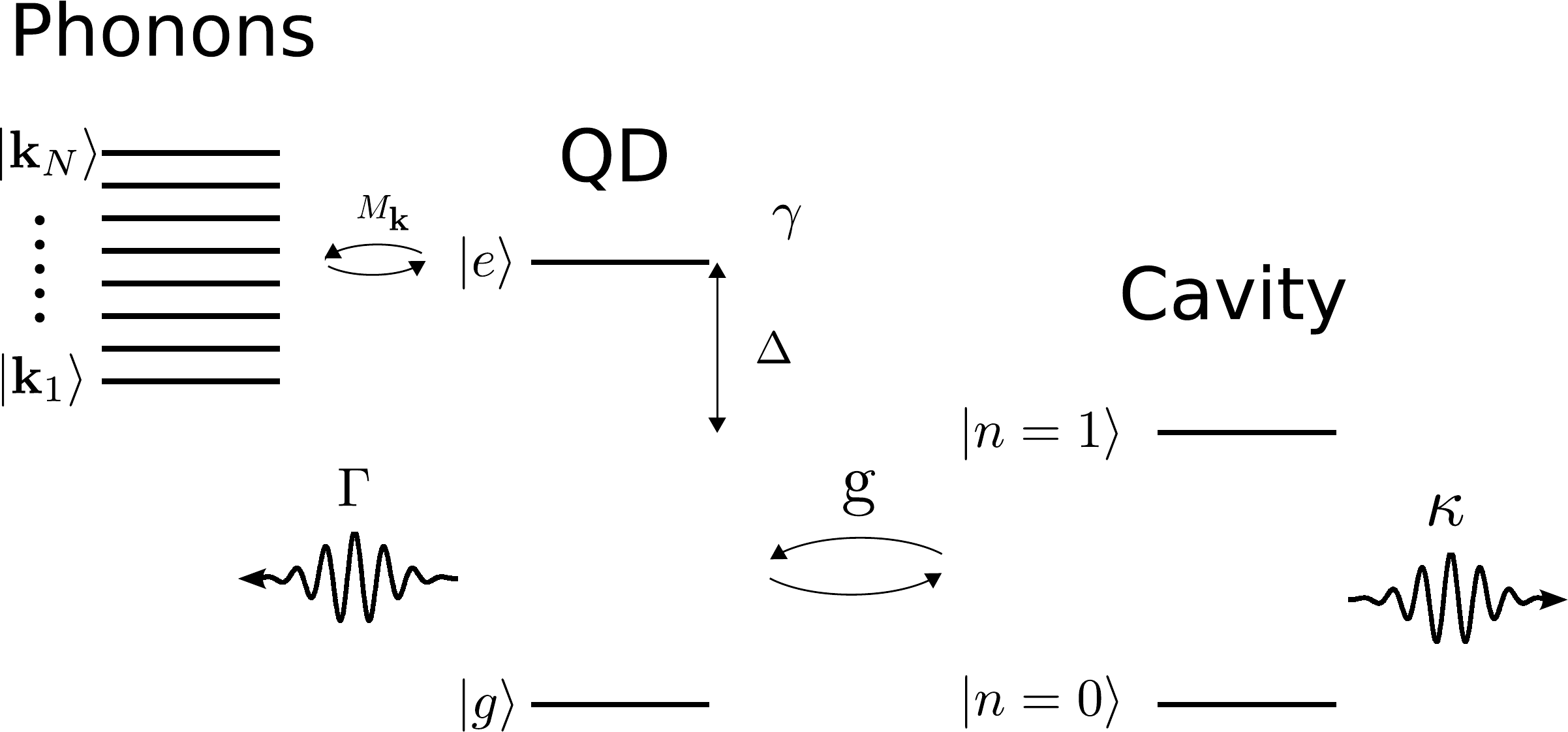}
 \caption{Schematic illustration of the cavity QED model system including the longitudinal acoustical phonon interaction. Important parameters include: The QD-cavity coupling strength $g$, the QD-cavity detuning $\Delta$, and the QD-phonon interaction matrix elements $M_{\bs{k}}$. The rates $\Gamma$ and $\kappa$ represent decay of the QD and cavity, respectively, and pure dephasing processes, beyond those induced by the direct phonon interaction, are included through the rate $\gamma$.
}
\label{fig:cQED_schematic}
\end{figure}

We model the QD-cavity system coupled to longitudinal acoustical (LA) phonons using a standard Hamiltonian
\cite{Wilson-Rae2002,Hohenester2009c,Hohenester2010,Kaer2010,Roy2011b,Kaer2012c,Kaer2013a,Kaer2013}, employing the following Hilbert space for the QD-cavity system: $\bT{\ket 1 = \ket{\mrm e, n=0},\ket 2 = \ket{\mrm g, n=1},\ket 3 = \ket{\mrm g, n=0}}$, where $n$ refers to the number of cavity photons and e (g) is the excited (ground) state of the QD.
Besides LA phonons, also longitudinal optical phonons couple strongly to QDs, however due to their large energy, on the order of tens of meV ($\sim 37~$meV for GaAs \cite{Krummheuer2002}), they are not important for the situations considered here, and are thus neglected in the model.
The system is illustrated schematically in \pref{fig:cQED_schematic}.

With reference to the total Hamiltonian of the system+reservoir, \pref{eq:uber_H}, we have
\al{\label{eq:Hs_def_first}
  H_\mrm S = \hb\Delta\sig_{11}+\hb g (\sig_{12}+\sig_{21}),
}
\al{\label{eq:H0_ph_def}
H_\mrm{R} &= \sum_{\bs k} \hb\om_\mrm{\bs k}\hc b_\mrm{\bs k}b_\mrm{\bs k},
}
\al{\label{eq:Heph_def_first}
  H_\mrm{I} = \sig_{11}\sum_{\bs k} M^\mrm{\bs k}(\hc b_\mrm{-\bs k}+b_\mrm{\bs k})=\sig_{11}R,
}
where $\sig_{nm}=\vert n\rangle \langle m \vert $ and $\hc b_\mrm{\bs k}$ and $b_\mrm{\bs k}$ are standard bosonic operators for the $\bs k$'th phonon mode.
We assume linear dispersion for the phonons, $\om_\mrm{\bs k}=c_\mrm s k$, with $c_\mrm s$ being the speed of sound.
Furthermore we introduced the effective matrix element \cite{Kaer2012c} taking into account the interaction with both the excited and ground state of the QD
\al{\label{eq:bare_phonon_ME}
M^{\bs k}= \sqrt{\frac{\hbar k}{2 d c_\mrm s V}} \int d\bs r\bS{ D_\mrm e|\phi_\mrm e(\bs r)|^2-D_\mrm g|\phi_\mrm g(\bs r)|^2} e^{-i\bs k \cd \bs r},
}
where $D_\nu$ are deformation potential constants, $d$ is the material mass density, $c_\mrm s$ is the speed of sound, $V$ is the phonon quantization volume, and $\phi_\nu(\bs r)$ are QD wavefunctions.
If, for simplicity, we assume an isotropic harmonic confinement for the QD potential, the wavefunctions are spherically symmetric
\al{
\phi_\nu(r)=\frac 1{\pi^{3/4}l^{3/2}_\nu}e^{-r^2/(2l^2_\nu)},
}
and the phonon matrix element becomes
\al{\label{eq:bare_phon_ME_spherical}
M^{k} = \sqrt{\frac{\hbar k}{2dc_\mrm{s}V}}\bS{D_\mrm{e}e^{-\frac 14(kl_\mrm e)^2}-D_\mrm{g}e^{-\frac 14(kl_\mrm g)^2}},
}
Using typical parameters for GaAs we have:
  $d=5370~\mrm{kg}\mrm{m}^{-3}$, $c_\mrm s=5110~\mrm{m}\mrm{s}^{-1}$, $D_\mrm e
  = -14.6~\mrm{eV}$, $D_\mrm g = -4.8~\mrm{eV}$, see e.g. Ref. \onlinecite{Krummheuer2002}.
The size of the QD wavefunction will be varied, but typically we use $l_\mrm \nu = 5~$nm.

We note that more realistic QD confinement potentials can be employed \cite{Nysteen2013}, but for our purposes isotropic harmonic confinement captures the most important features.

Having specified the phonon interaction in detail, we can now calculate the relevant correlation function describing the properties of the phonon reservoir, which we assume to be in a thermal state.
From \pref{eq:uber_D_def} we have
\al{\label{eq:Dph_def}
D(t)&=\sum_{\bs k} \abs{M^{k}}^2 \bS{n_{k}\e{+ i \omm{k}t}+\bP{n_{k}+1}\e{- i \omm{k}t}}\\
&=\sum_{\bs k} \abs{M^{k}}^2 \bS{(2n_{k}+1)\cos (\omm{k}t) - i \sin (\omm{k}t)},
}
where $n_k$ is the thermal occupation of the $\bs k$'th phonon mode, given by the Bose-Einstein factor $n_k=1/(\exp [\hbar \om_k/(k_\mrm B T)]-1)$.
We now define a spectral function of the phonon reservoir as the real part of the Fourier transform of $D(t)$
\al{\label{eq:eff_phonon_density}
d_\mathrm{ph}(\om)=\pi\sum_{\bs k}\abs{{M}_{k}}^{2}\bS{(n_{k}+1)\delta(\om-\om_{k})+n_{k}\delta(\om+\om_{k})}.
}
As we shall see, $d_\mathrm{ph}(\om)$ can also be interpreted as an effective phonon density and is a very useful concept in understanding the properties of the cQED system.
Terms proportional to $(n_k+1)$ are responsible for phonon emission processes for $\om>0$, where the ``$1$" signifies spontaneous phonon emission through stimulation by the ever-present phonon vacuum field.
The presence of the phonon vacuum field has the consequence that it is impossible to completely ``freeze out" phonon processes, as these are present even at $T=0$ where $n_k=0$.
Terms proportional to $n_k$ are responsible for phonon absorption processes for $\om<0$.

\begin{figure}[ht]
 \includegraphics[width=0.48\textwidth]{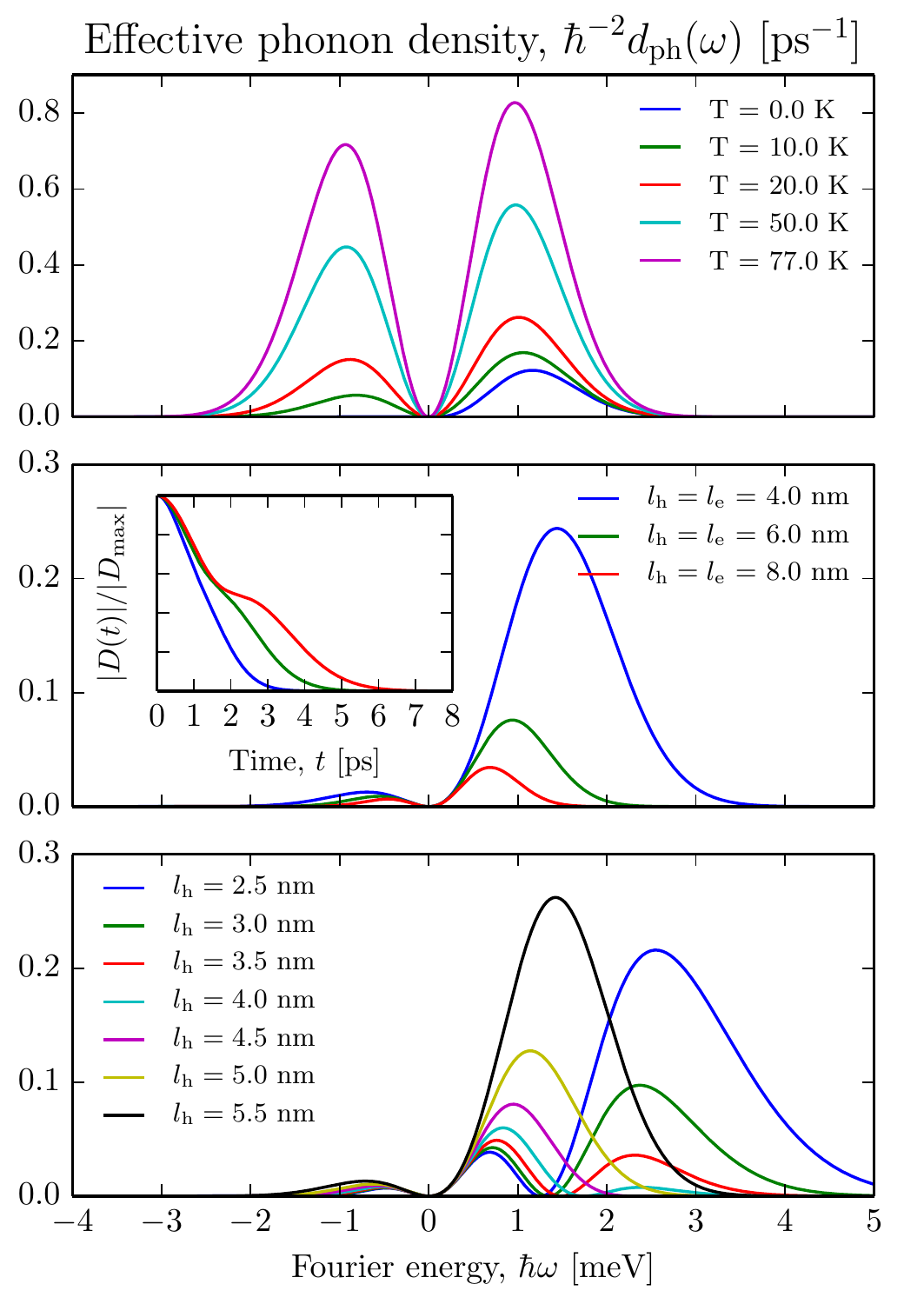}
 \caption{Effective phonon spectrum  $d_\mathrm{ph}(\om)$, given by \pref{eq:eff_phonon_density_expli}. (Top) Temperature is varied while the QD confinement lengths are fixed $l_\mrm h = l_\mrm e=5~$nm. (Middle) Temperature is fixed $T=4~$K, while $l=l_\mrm h = l_\mrm e$ is varied. Inset: The corresponding phonon correlation function, \pref{eq:Dph_def}. (Bottom) Temperature is fixed $T=4~$K and the QD confinement lengths are varied while keeping $l^3_\mrm h + l^3_\mrm e=2\times(5~\mrm{nm})^3$ fixed.}
\label{fig:Dph_plots}
\end{figure}

The effective phonon density has a direct relation to Purcell enhanced QD lifetimes \cite{Kaer2012c} and has recently been measured \cite{Madsen2013}.
Inserting \pref{eq:bare_phon_ME_spherical} in to \pref{eq:eff_phonon_density} we obtain an explicit expression for $d_\mathrm{ph}(\om)$
\ml{\label{eq:eff_phonon_density_expli}
d_\mathrm{ph}(\om)=\frac{\hbar}{4 \pi d c^5_\mrm{s}}\frac{\om^3}{1-e^{-\hbar \om/(k_\mrm B T)}}\\
\times\bS{D_\mrm{e}e^{-\frac 14(\om l_\mrm e/c_\mrm s)^2}-D_\mrm{g}e^{-\frac 14(\om l_\mrm g/c_\mrm s)^2}}^2.
}
As an illustration, \pref{fig:Dph_plots} shows the effective phonon density $d_\mathrm{ph}(\om)$ for different values of the QD confinement lengths, $l_\mrm e$ and $l_\mrm g$, and temperature.
The main feature of $d_\mathrm{ph}(\om)$ is a pronounced peaked structure with a maximum typically in the $\abs{\hbar\om}\sim 1- 2~$meV range, except for the bottom figure where several maxima appear \cite{Nysteen2013}. The exact position of the maximum is determined by a combination of the QD confinement lengths.
In the special case of $T=0$ and $l=l_\mrm e=l_\mrm g$ the maximum is located at $\om_\mrm{max}=\sqrt{3}c_\mrm s/l$.

We include the decay of photons escaping from the cavity at a rate $\kappa$, related to the $Q$-factor as $Q=\om_\mrm{cav}/\kappa$, causing the system to make a transition from $\vert \mrm g, 1\rangle$ to $\vert \mrm g, 0\rangle$, by including a Lindblad term $\alpha=\kappa$ and $P=\vert \mrm g, 0\rangle \langle \mrm g, 1\vert= \sigg{32}$ with reference to \pref{eq:twotime_Lindblad}.
Decay of the QD, into optical modes other than the cavity and due to non-radiative channels, is similarly included through a Lindblad term $\alpha=\Gamma$ and $P=\vert \mrm g, 0\rangle \langle \mrm e, 0\vert=\sigg{31}$.
For completeness, we also include pure dephasing processes beyond those induced by the LA phonon reservoir using the Lindblad term $\alpha=2\gamma$ and $P=\vert \mrm e, 0\rangle \langle \mrm e, 0\vert=\sigg{11}$, although in the results to be presented we always take $\gamma=0$.
These additional dephasing processes could arise from the phonon-induced coupling to higher shells in the QD \cite{Muljarov2004} or random nuclear field flutuations \cite{Greilich2006}.

The full set of dynamical equations are presented in their entirety in \pref{app:EOM_RDM}.
Both the one-time and two-time equations are linear systems of ordinary differential equations with time-dependent coefficients, which means that they can easily be solved using standard software packages.
For further discussions of approximations we refer to \pref{sec:useful_approximations}.
\section{Results}\label{sec:results}
In this section we present results for the degree of indistinguishability in dependence on important parameters for the cQED system.
The indistinguishability is related to the normalized number of coincidence events \cite{Kiraz2004} in a Hong-Ou-Mandel (HOM) experiment \cite{Hong1987} and is defined as \cite{Kiraz2004}
\al{
I&=1-\frac{\int_{-\infty}^{+\infty}dt\int_{-\infty}^{+\infty}d\tau G^{(2)}_{\mrm{HOM}}(t+\tau,t)}{\int_{-\infty}^{+\infty}dt\int_{-\infty}^{+\infty}d\tau G^{(2)}_{\mrm{uncorr}}(t+\tau,t)}\\\label{eq:ID_def}
&=\frac{\int_{0}^{\infty}dt\int_{0}^{\infty}d\tau\abs{\langle \hc A(t+\tau)A(t)\rangle}^{2}}
{\int_{0}^{\infty}dt\int_{0}^{\infty}d\tau\langle \hc A(t+\tau)A(t+\tau)\rangle
\langle \hc A(t)A(t)\rangle}.
}
Here $G^{(2)}_{\mrm{HOM}}(t+\tau,t)$ is the second order correlation function for the HOM experiment and $G^{(2)}_{\mrm{uncorr}}(t+\tau,t)$ accounts for the uncorrelated coincidence events, which can be modeled as an HOM experiment without beam-splitter.
The operator $A$ is chosen corresponding to the relevant photon quantum field.
For the case of cavity emission, $A$ equals the cavity photon annihilation operator, $A=\vert \mrm g, 0\rangle \langle \mrm g, 1\vert=\sigg{32}$, corresponding to energy leaving the system through the $\kappa$ decay channel.
For QD emission, $A$ equals the QD de-excitation operator, $A=\vert \mrm g, 0\rangle \langle \mrm e, 0\vert=\sigg{31}$, corresponding to energy leaving the system through the $\Gamma$ decay channel.

For all results presented here, the QD is initially excited, the cavity is in its ground state, and the phonons are in a thermal state determined by the given temperature.
This initial condition simulates pulsed resonant excitation of the QD as well as non-resonant excitation in the case where relaxation processes populate the excited QD state faster than all other timescales in the system.

While cavity and QD emission are easily separated formally, the experimental distinction between these emission channels is non-trivial.
To separate the light into a cavity and QD part, either spectral and/or spatial selection methods may be used.
Spectral separation is complicated by the fact that QD and cavity emission often overlap spectrally and might even be strongly coupled, which causes the two signals to overlap and mix in complicated ways.
Spatial separation depends on the specific optical structure employed to form the cavity.
In a micropillar cavity structure, spatial separation of the QD and cavity emission is often possible, due to the directed nature of the cavity itself.
Here it is known that light emitted through the top mirrors, along the growth axis, is strongly dominated by cavity emission, whereas QD emission mostly couple to modes other than the strongly directional cavity mode \cite{Ates2009b}.
Due to the complicated geometry of a photonic crystal it is notoriously difficult \cite{Madsen2013a} to assign a specific spatial direction to light emitted from either QD or cavity.

Another important figure of merit for a single photon source is the fraction of light emitted into the cavity mode of interest, denoted the cavity emission efficiency and defined as
\al{\label{eq:beta_cav}
\beta_\mrm{cav}=\frac{\kappa\int dt n_\mrm{cav}(t)}{\kappa\int dt n_\mrm{cav}(t)+\Gamma\int dt n_\mrm{QD}(t)},
}
where $n_\mrm{cav}(t)$ and $n_\mrm{QD}(t)$ are the populations of the cavity mode and QD, respectively.
This expression is valid in the weak as well as the strong coupling regime and in the weak coupling regime it reduces to the usual expression in terms of the Purcell factor \cite{Kaer2013a}.

\subsection{Validity of perturbational treatment of phenomenological decay in conjunction with non-Markovian reservoirs}
In the regime of strong coupling  between light and matter, the initially excited QD is strongly affected by the interaction with the cavity.
However, as the losses increase, in practice typically the cavity decay rate $\kappa$, the system enters the weak coupling regime, where the main role of the cavity is reduced to providing an additional decay channel, giving rise to the well-known Purcell enhancement, through an additional rate $\Gamma_\mrm{Purcell}$, adding to the background decay rate $\Gamma$.
Further increasing $\kappa$, the cavity becomes less and less important and in the limit of $\kappa\rightarrow\infty$ we have $\Gamma_\mrm{Purcell}\rightarrow 0$, and the cavity will simply act as a very weak filter of the light emitted from the cavity, not changing its basic properties, only diminishing the intensity of light exiting via the cavity decay channel.

In this limit of very low quality cavities we therefore expect the degree of indistinguishability for QD and cavity to approach each other, $\lim_{\kappa\rightarrow\infty}(I_\mrm{QD}-I_\mrm{cav})=0$.
Formally, we can extrapolate to this limit by setting $g=0$ in our model, which decouples the cavity from the QD, and \pref{eq:super_uber_twotime_EOM} can be solved in closed form\cite{Kaer2013,Kaer2013a,Nazir2009b} yielding
\al{\label{eq:I_QD_g0}
I_{g=0}=\Gamma\int_0^{\infty}d\tau \exp (-\Gamma\tau-2\mrm{Re}[\varphi(0)-\varphi(\tau)]),
}
where $\varphi(\tau)=-\hbar^{-2}\int d\tau\int d\tau D(\tau)$ and $D(\tau)$ is the phonon reservoir correlation function, \pref{eq:Dph_def}.
This function is well-known from polaron transformation approaches \cite{Kaer2012c}.
In fact, for $g=0$ our model basically reduces to the exactly solvable independent boson model, except that we include a small loss rate, for which reason we expect \pref{eq:I_QD_g0} to be accurate even for high temperatures and large phonon coupling strengths.

\begin{figure}[ht]
 \includegraphics[width=0.48\textwidth]{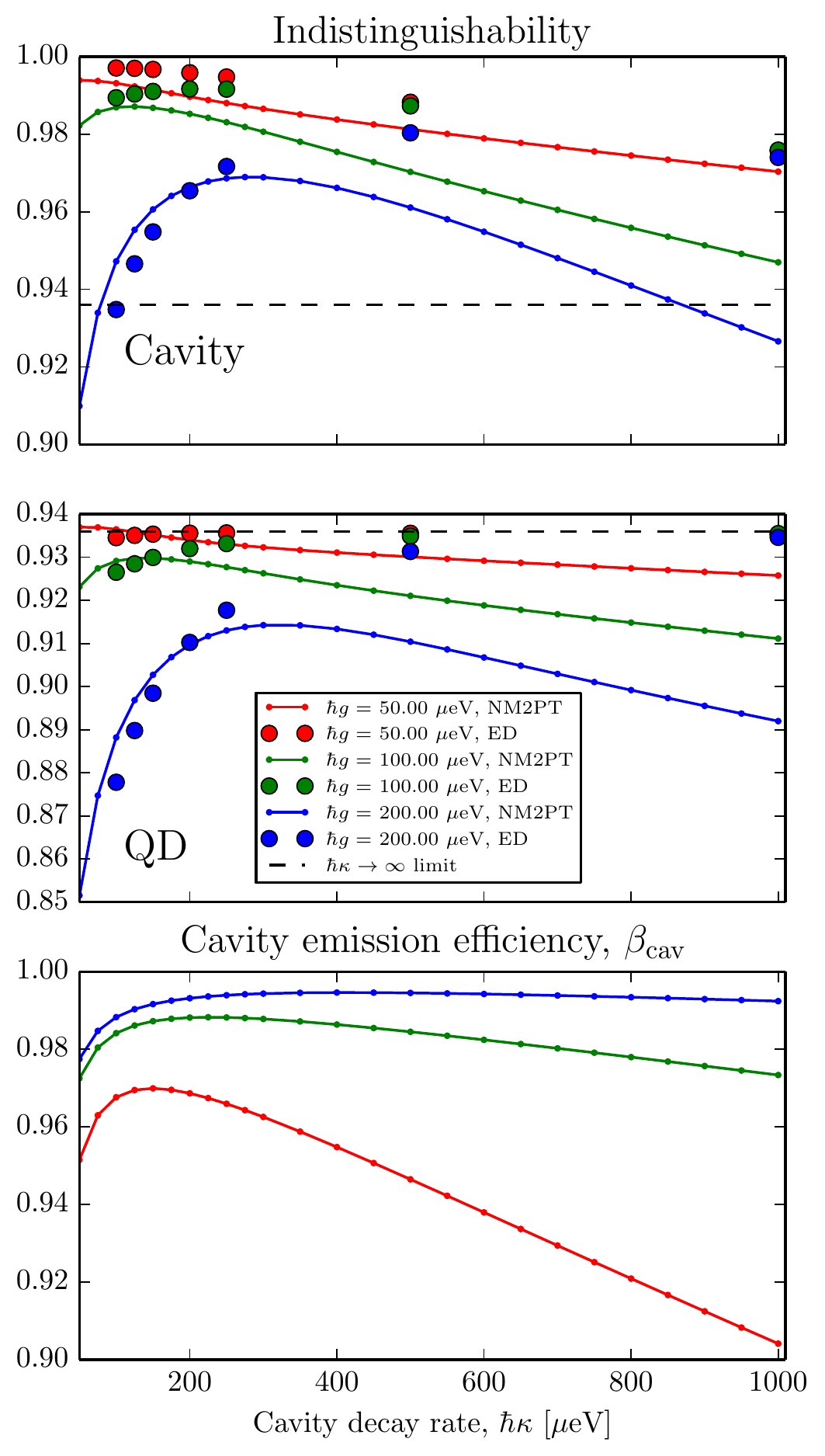}
 \caption{Indistinguishability (two upper plots) as a function of the cavity decay rate, calculated from \pref{eq:super_uber_twotime_EOM} and using an exact diagonalization (ED) approach \cite{Hohenester2007,Kaer2013,Kaer2013a} and cavity emission efficiency $\beta_\mrm{cav}$, see \pref{eq:beta_cav} (lower plot). Parameters: $\hbar \Gamma=1~\mu$eV, $\hbar \Delta=0$, $T=0~$K, and $l_\mrm h = l_\mrm e=5~$nm.}
\label{fig:U_decay_IIII_compare_ED}
\end{figure}
In \pref{fig:U_decay_IIII_compare_ED} we show the indistinguishability as a function of the cavity decay rate $\kappa$ for a few representative values of the QD-cavity coupling $g$, to cover both weak and strong coupling, along with the $\kappa\rightarrow\infty$ result of \pref{eq:I_QD_g0}.
In the present context, the most interesting behavior occurs for $\hbar\kappa> 300~\mu$eV, where the curves for the NM2PT do not seem to converge towards the large $\kappa$ result, indicated by the dashed line, as would have been expected.
The departure from the $\kappa\rightarrow\infty$ results continues at least until $\hbar\kappa=4000~\mu$eV (not shown).
For comparison we have also included the indistinguishability obtained using an ED approach \cite{Hohenester2007,Kaer2013,Kaer2013a}, which treats the phonon interaction without any approximations except for a controlled truncation of the phonon Hilbert space.
The two methods show fairly good agreement up until $\hbar\kappa\sim 300~\mu$eV (the error on the ED results is approximately $0.005$), after which only the ED result seems to converge toward the expected large $\kappa$ result.
The convergence is almost complete for QD emission, whereas for cavity emission it is significantly slower. However, we have performed (less accurate) simulations up to $\hbar\kappa=4000~\mu$eV, where the cavity is within $0.01$ of having converged to the large $\kappa$ value.

From the comparison with the large $\kappa$ result and the ED approach, it is clear that the NM2PT breaks down for large cavity decay rates.
The deviations probably arise from the assumption of uncorrelated and independent reservoirs, which is used in the derivation of \pref{eq:super_uber_twotime_EOM} and also the Lindblad formula \pref{eq:twotime_Lindblad}.
This has the consequence that scattering rates from different reservoirs can simply be added independently, in the Lindblad formalism this is expected to be an extremely good approximation due to the absence of memory in the reservoir.
For the usual Markovian situation, this does not cause a problem, however this is not the case for a spectrally structured non-Markovian reservoir.
From a more practical point of view, an immediate problem is that the non-Markovian scattering rates in \pref{eq:super_uber_twotime_EOM} only contain the unitary properties of the small system, more specifically the interaction picture time evolution of the $P$ operators is only governed by the coherent Jaynes-Cummings Hamiltonian $H_\mrm{JC}$, see \pref{eq:Hs_def_first} and does not contain any loss.
Effectively, this means that the phonons will always interact with a non-decaying polariton quasi-particle, and not the real lossy quasi-particle that might have completely different properties.
This feature is a consequence of the perturbational nature of our theory, which assumes that the small system only interacts weakly with a reservoir and therefore the associated rate is small compared to other system inverse timescales.
This is well fulfilled for small cavity decay rates $\kappa$, but the approximation breaks down for large $\kappa$.
Again, usually this is not a problem, as Markovian reservoirs do not ``see" any other reservoirs due to their infinitely short memory.
The reason why the ED approach gives the correct result is it that treats both electrons, photons, and phonons on equal footing, thus the phonons will interact with a QD that is dressed by a lossy cavity rather than the perfect polariton assumed by our present theory.

\begin{figure}[ht]
 \includegraphics[width=0.48\textwidth]{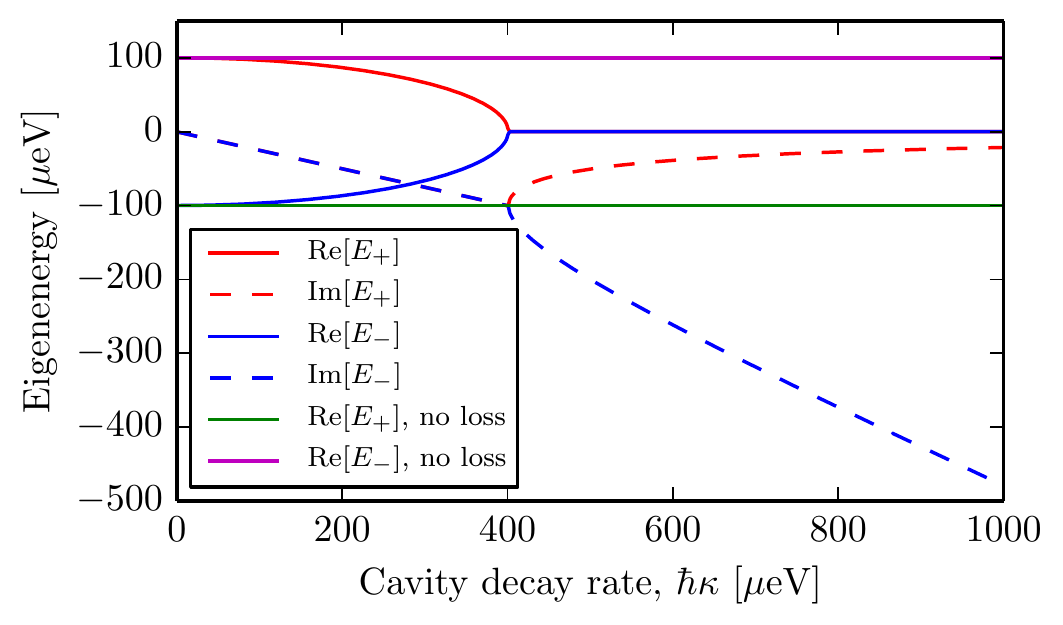}
 \caption{Illustration of the eigenenergies of the Jaynes-Cummings Hamiltonian with and without loss, \pref{eq:lossy_EVs}. Note that the blue and red curves overlap in parts of the figure. Parameters: $\hbar \Gamma=1~\mu$eV, $\hbar g=100~\mu$eV, $\hbar \Delta=0$.}
\label{fig:H_evals}
\end{figure}
Inspired by this observation, the situation might be remedied by including loss in the Jaynes-Cummings Hamiltonian responsible for the interaction picture time evolution in the scattering terms of \pref{eq:super_uber_twotime_EOM}, i.e. instead of \pref{eq:Hs_def_first} we use the following
\al{\label{eq:lossy_HS}
H_\mrm{JC}/\hbar\rightarrow\bmm{
\Delta-i\frac\Gamma 2 & g\\
g & -i\frac\kappa 2
}.
}
The eigenenergies of this matrix are
\al{\label{eq:lossy_EVs}
E_\pm=\frac\hbar 2\bS{\Delta-i\frac{\Gamma+\kappa}{2}\pm\sqrt{4g^2-\bS{\frac\Gamma 2-\frac\kappa 2+i\Delta}^2}},
}
and are plotted in \pref{fig:H_evals} versus cavity decay rate
The figure shows that loss strongly affects the spectral properties of the Jaynes-Cummings system, both through the spectral position (real part) and decay (imaginary part) of the eigensolutions.
However, the replacement of $H_\mrm S$ with the phenomenological \pref{eq:lossy_HS} was found not to give a systematic improvement of the predicted degree of indistinguishability compared to the ED.
Furthermore, simply replacing the Jaynes-Cummings Hamiltonian with its non-Hermitian version \emph{after} the derivation, has the consequence that the time-evolution operator $U(t)$ becomes non-unitary, which violates the basic assumptions of the theory and the end result will depend on exactly when in the derivation the substitution was made.
A proper inclusion of the effect of loss in the phonon scattering terms of the present perturbational theory, requires one to go back and closely examine the approximations made in the derivation, especially concerning the correlations between the different reservoirs included in the model. This endeavour is, however, beyond the scope of the present work.

\begin{figure}[ht]
 \includegraphics[width=0.48\textwidth]{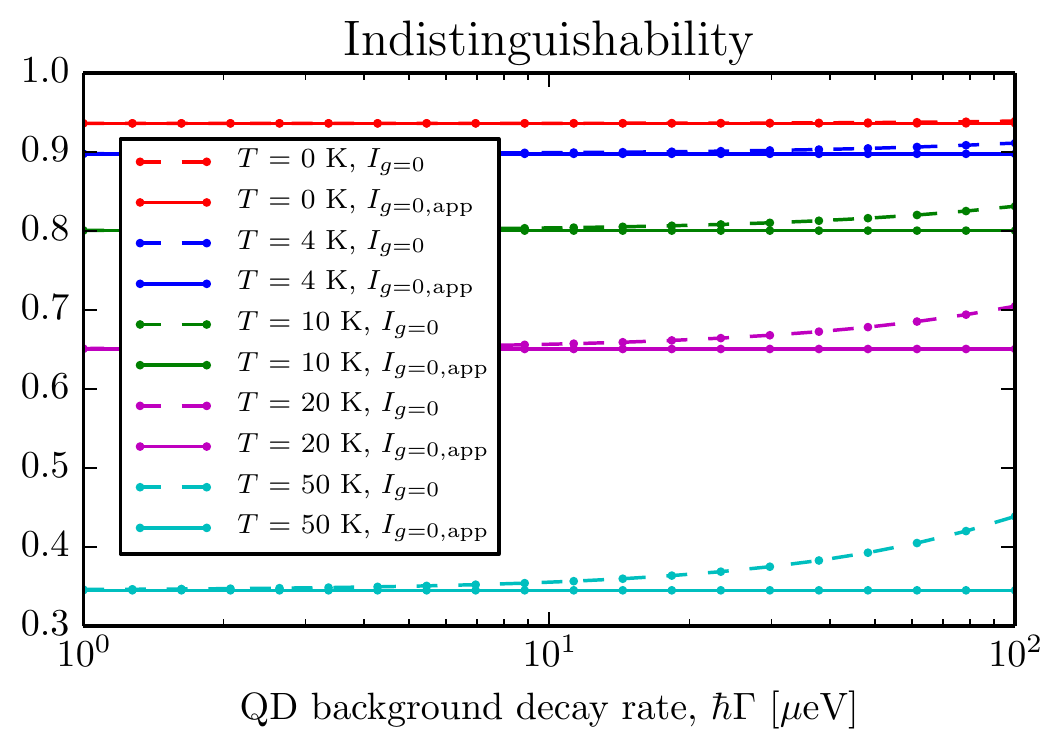}
 \caption{Indistinguishability for $g=0$, where $I_{g=0}$ is from \pref{eq:I_QD_g0} and $I_{g=0,\mrm{app}}$ is from \pref{eq:I_QD_g0_app}. Parameters: $l_\mrm h = l_\mrm e=5~$nm.}
\label{fig:ID_WC}
\end{figure}
The results of \pref{fig:U_decay_IIII_compare_ED} limits the range of $\kappa$ values that can be investigated with the NM2PT.
Closely examining the different curves, nevertheless show a clear tendency for larger $g$ simulations to remain accurate up to larger $\kappa$ values.
Intuitively this makes sense, as our basic theory assumes cavity decay to constitute a perturbation to the coherent system dynamics, leading to the requirement $\kappa < g$.
For the remainder of the paper we have chosen parameter values for which the NM2PT should yield correct results.

As noted above, the indistinguishability for the QD emission quickly converges to the large $\kappa$ limit result, where \pref{eq:I_QD_g0} applies and provides a much faster method for calculating the indistinguishability than solving dynamical two-time EOMs.
However, it should be noted that for these values of $\kappa$ the cavity is still the dominant emission channel, as shown by the $\beta_\mrm{cav}$ calculations in the bottom panel og \pref{fig:U_decay_IIII_compare_ED}.

Figure \ref{fig:ID_WC} shows the indistinguishability calculated from \pref{eq:I_QD_g0} as a function of the QD background decay rate $\Gamma$, for a range different temperatures.
The figure shows that for realistic values of the QD background decay rate, typically on the order of $\hbar\Gamma\sim 1\mu$eV corresponding QD lifetime in the 500 to 1000 ps range, the indistinguishability depends very weakly on $\Gamma$, only for unrealistically large $\Gamma$ values do we observe a significant dependence.
We note that the effective QD decay rate can become large, i.e. through the Purcell effect, but in this case the cavity is part of the system and \pref{eq:I_QD_g0} no longer applies.
We can approximate \pref{eq:I_QD_g0} as follows
\al{\notag
I_{g=0}&=\mrm e^{-2\mrm{Re}[\varphi(0)]}\Gamma\int_0^{\infty}d\tau \mrm{e}^{-\Gamma\tau} \mrm e^{2\mrm{Re}[\varphi(\tau)]}\\\notag
&\approx\mrm e^{-2\mrm{Re}[\varphi(0)]}\Gamma\int_0^{\infty}d\tau \mrm{e}^{-\Gamma\tau}(1+2\mrm{Re}[\varphi(\tau)])\\\notag
&=\mrm e^{-2\mrm{Re}[\varphi(0)]}(1+\Gamma\int_0^{\infty}d\tau \mrm{e}^{-\Gamma\tau}2\mrm{Re}[\varphi(\tau)]).
}
In the second line we Taylor expanded the exponential containing $\varphi(\tau)$, which is an excellent approximation especially at low temperatures, see Fig. 2 of Ref. \onlinecite{Kaer2012c}.
The third line shows that the contribution from the last integral can be thought of as frequency filtering of $\varphi(\tau)$ near $\om=0$ and with a bandwidth of $\Gamma$, and as $\varphi(\tau)$ has a frequency spectrum similar to the spectrum of $D(t)$, see \pref{fig:Dph_plots}, we expect this contribution to be small for realistic $\Gamma$ values.
We get
\al{\label{eq:I_QD_g0_app}
I_{g=0,\mrm{app}}=\mrm e^{-2\mrm{Re}[\varphi(0)]},
}
which is shown in \pref{fig:ID_WC} together with $I_{g=0}$ and we see that the approximations made above are well justified.
The simple result in \pref{eq:I_QD_g0_app} has several important consequences.
First, it provides a simple way of calculating the indistinguishability for a QD interacting with LA phonons, as one only needs to evaluate the simple integral
\al{
\varphi(0)=\sum_{\bs k} \abs{M^{k}/(\hbar\om_{k})}^2 \bS{2n_{k}+1}.
}
Second, it shows that the indistinguishability does not depend on the QD lifetime, as long as the lifetime is long compared to the phonon reservoir correlation time.
Third, the QD filters out the phonon sidebands, evidenced by the fact that \pref{eq:I_QD_g0_app} does not depend on the time-dependent $\varphi(\tau)$, only on its value at zero time.
\subsection{Time-dependent pure dephasing rates}\label{sec:puredephrates}
Important insights about the phonon-induced decoherence may be obtained by inspection of the different terms in the EOMs and their time-dependence.
The time-dependence arises from the non-Markovian nature of the phonon reservoir and is especially important to consider as the degree of indistinguishability is derived from a two-time function, where the time-dependence of the scattering rates persist even in the long-time limit, see Ref. \onlinecite{Kaer2013} for further discussion and illustrations.
The most important terms are those giving rise to pure dephasing effects.
The pure dephasing rate for the cavity emission is, see \pref{eq:twotime_M22},
\al{\label{eq:M22_cavity}
M_{2,2}(t,\tau)\lvert_\mrm{ph}&=-H^*_{11,11}(\tau+t,0)+G_{12,12}(t+\tau,\tau),
}
and for QD emission, see \pref{eq:twotime_M33},
\al{\label{eq:M33_QD}
M_{3,3}(t,\tau)\lvert_\mrm{ph}&=-H^*_{11,11}(\tau+t,0)+G_{11,11}(t+\tau,\tau).
}
Note that these rates enter in the EOMs in the difference time $\tau$, e.g. $\partial_{\tau}\braket{a^\dagger(t+\tau)a(t)}$, and thus the absolute time $t$ is a fixed parameter.
\begin{figure}[ht]
 \includegraphics[width=0.48\textwidth]{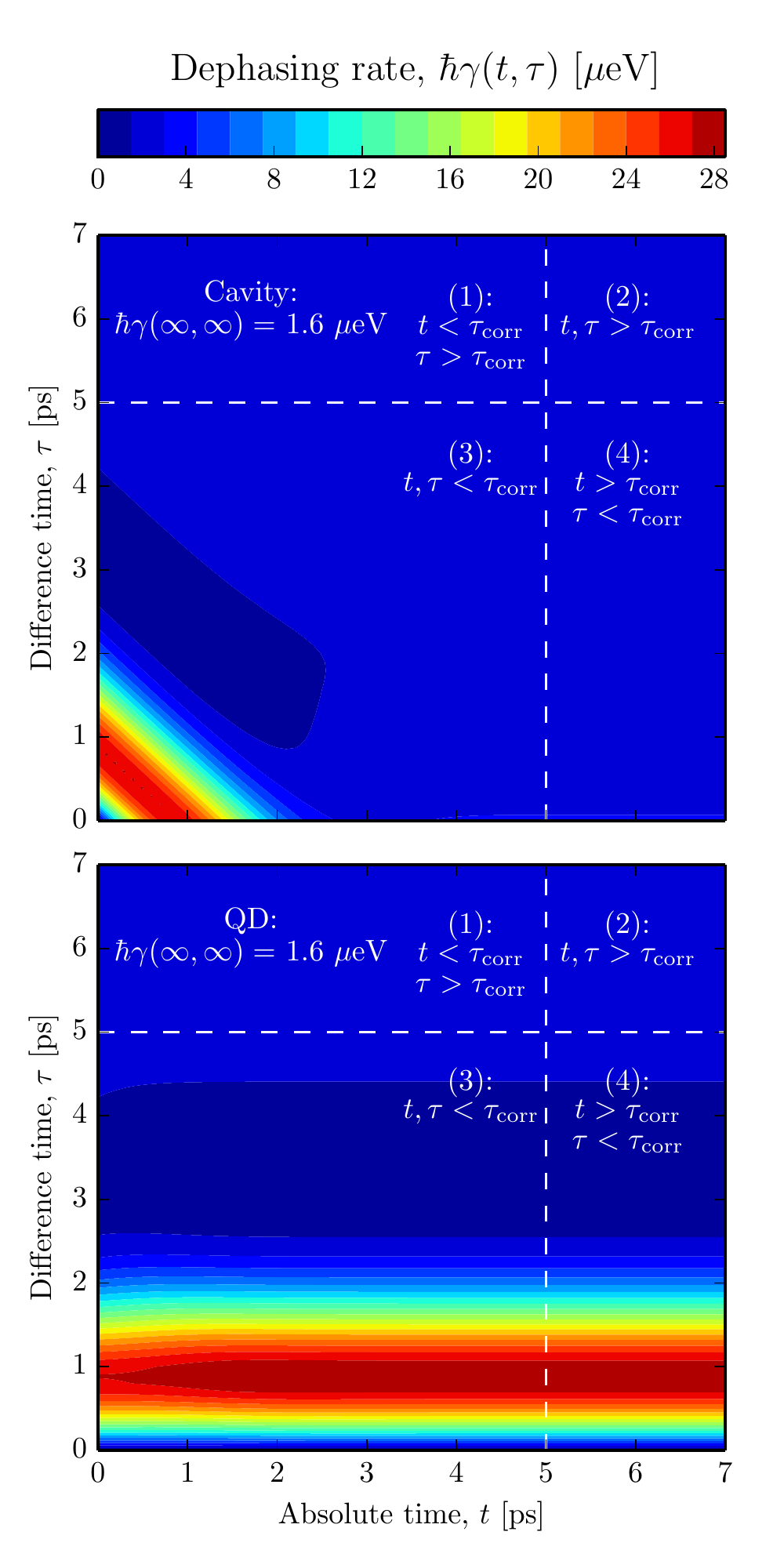}
 \caption{Real part of time-dependent two-time dephasing rates for QD, \pref{eq:M33_QD}, and cavity, \pref{eq:M22_cavity}. Parameters: $\hbar g=100~\mu$eV, $\hbar \Delta=0$, $T=4~$K, and $l_\mrm h = l_\mrm e=5~$nm.}
\label{fig:twotime_3d_g100}
\end{figure}

In \pref{fig:twotime_3d_g100} we show the real part of these rates as a function of $\tau$ and $t$ for a QD-cavity coupling of $\hbar g=100~\mu$eV. Note that the phonon induced scattering rates do not depend on the relaxation rates of the system, see the discussion in the previous section on large $\kappa$ behavior.
Also, for $t=0$ the QD and cavity rates are identical and show the well-known time-dependent behavior of the short-time non-Markovian regime \cite{Kaer2013}, with large values initially and a smaller constant value for longer times.
The important timescale here is the decay time of the reservoir correlation function, $D(t)$, which for the typical parameters used here is around 5 ps, see inset in \pref{fig:Dph_plots} for more examples.
This characteristic decay time sets the timescale over which the rates remain time-dependent and is often referred to as the memory depth of the reservoir or the reservoir correlation time $\tau_\mrm{corr}$.
At $t=1~$ps, the two rates depend quite differently on $\tau$ and for $t>\tau_\mrm{corr}$ the rates no longer depend on $t$ and show qualitatively different behavior: The rate associated with the cavity basically becomes a constant, whereas the rate associated with the QD retains a strong $\tau$ time-dependence.
From the above observations it appears that the $\tau$-dependence of the rate relevant for QD emission does not change much as a function of $t$ and is rather well approximated by
\al{\label{eq:M33_cavity_approx}
M_{3,3}(t,\tau)\lvert_\mrm{ph}\approx-H^*_{11,11}(\tau,0);
}
The rate relevant for cavity emission depends strongly on $t$ for $t<\tau_\mrm{corr}$, but is for $t>\tau_\mrm{corr}$ well approximated by
\al{\label{eq:M22_cavity_approx}
M_{2,2}(t,\tau)\lvert_\mrm{ph}\approx-H^*_{11,11}(\infty,0),
}
i.e. it does not depend on neither $t$ nor $\tau$.
Combining \pref{eq:H_coef_def} and \pref{eq:U_op_explicit} we find that
\ml{\label{eq:longtime_M22_coef}
\mrm{Re}\bS{M_{2,2}(t,\tau)\lvert_\mrm{ph}}\propto d_\mrm{ph}(\om=\sqrt{4g^2+\Delta^2})\\
+d_\mrm{ph}(\om=-\sqrt{4g^2+\Delta^2}),
}
indicating that the effect of the phonons is effectively Markovian for cavity emission, in the sense that the scattering terms sample only distinct energies of the reservoir.
This is not the case for the QD emission, where the Markovian regime for the two-time scattering rates is only reached for $\tau>\tau_\mrm{corr}$, and a short-time non-Markovian regime is always present for $\tau<\tau_\mrm{corr}$, regardless of the value of the absolute time $t$.
In the non-Markovian regime, the entire phonon reservoir spectrum is sampled, leading to large dephasing rates, corresponding to virtual processes dominating the dephasing dynamics.

\begin{figure}[ht]
 \includegraphics[width=0.48\textwidth]{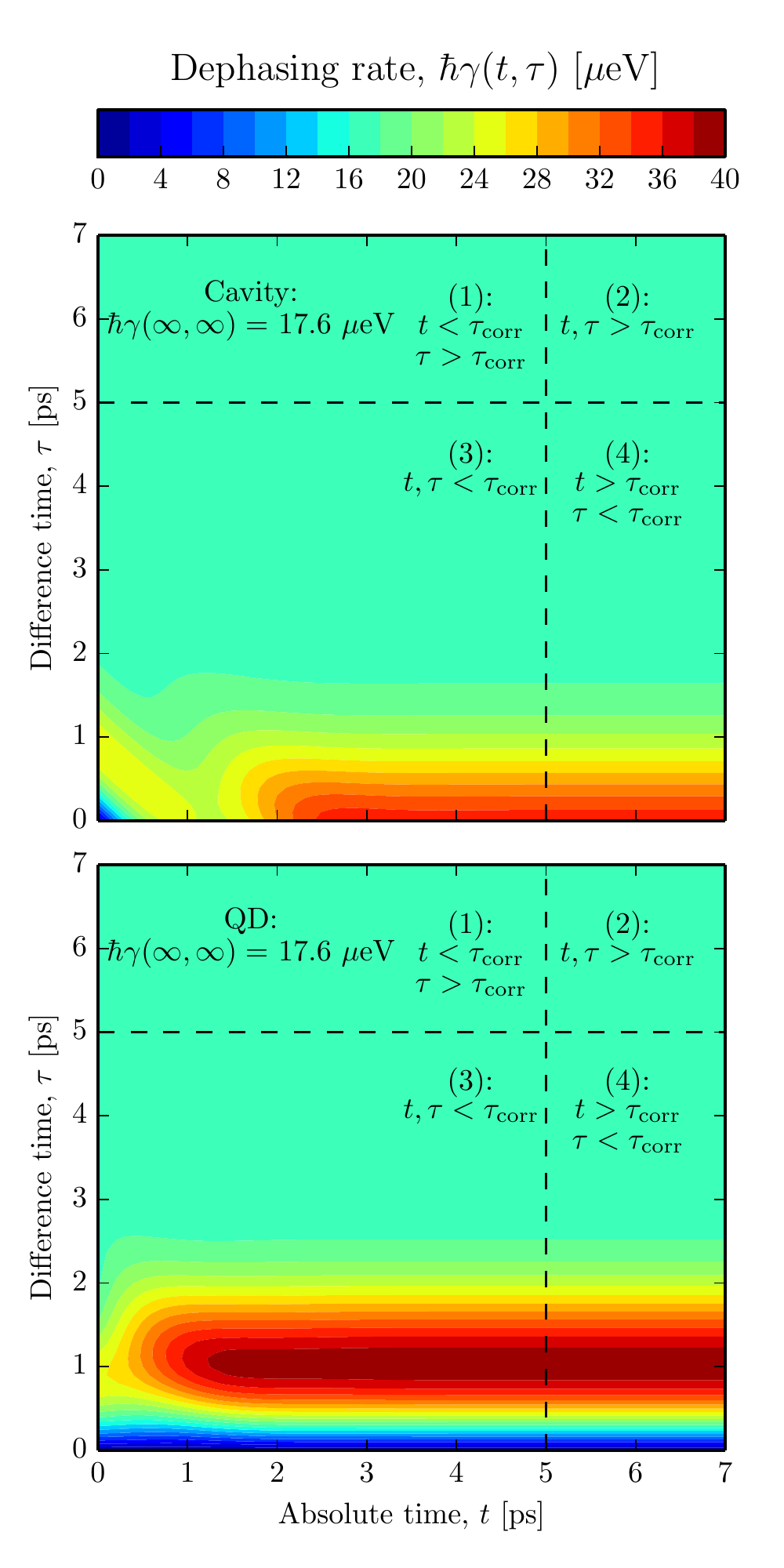}
 \caption{As \pref{fig:twotime_3d_g100} but with $\hbar g=400~\mu$eV.}
\label{fig:twotime_3d_g400}
\end{figure}
In \pref{fig:twotime_3d_g400} we increased the QD-cavity coupling to $\hbar g=400~\mu$eV.
While the QD dephasing rate only changes in overall amplitude, the cavity dephasing rate changes in a qualitative way.
Instead of basically displaying Markovian behavior for $t>3~$ps, as was the case for $\hbar g=100~\mu$eV, the rate maintains its time-dependence along the $\tau$ direction for all values of $t$ and thus its non-Markovian nature.
This means that the approximation for the cavity dephasing expressed in \pref{eq:M22_cavity_approx} no longer applies.
To summarize the different regimes of Markovian and non-Markovian behavior, we have divided the two-time plane into four regions depending on the value of $t$ and $\tau$ compared to $\tau_\mrm{corr}$, see \pref{tab:twotime_regimes} for an overview.
\begin{table}
\caption{\label{tab:twotime_regimes} Summary of the regimes of Markovian (M) and non-Markovian (nM) behavior in Figs. (\ref{fig:twotime_3d_g100}) and (\ref{fig:twotime_3d_g400}).}
\begin{ruledtabular}
\begin{tabular}{lllll}
Region & (1) & (2) & (3) & (4)\\\hline
$\hbar g=100~\mu$eV, cavity & M & M & nM & M\\
$\hbar g=100~\mu$eV, QD & M & M & nM & nM\\
$\hbar g=400~\mu$eV, cavity & M & M & nM & nM\\
$\hbar g=400~\mu$eV, QD & M & M & nM & nM\\
\end{tabular}
\end{ruledtabular}
\end{table}

The different behaviors of the cavity dephasing rate for small and large QD-cavity couplings can be qualitatively understood in the following way: The phonons interact directly with the QD degrees of freedom and  whatever dynamics they exhibit is thus directly transferred to the QD.
There is therefore no qualitative difference between the case of small and large values of $g$.
The cavity degrees of freedom, on the other hand, do not interact directly with the phonons and any interaction is mediated by the QD-cavity coupling.
The kind of dynamical response the QD-cavity coupling can mediate is hence determined by the magnitude of the QD-cavity coupling rate $g$,  with small $g$ values only mediating slow dynamics, since faster dynamics is filtered out, while larger $g$ values allow the ``transfer" of faster dynamics.
The characteristic timescale of the phonons is set by the correlation time $\tau_\mrm{corr}\sim 5~$ps, which should be compared to a typical timescale of the QD-cavity system, e.g. the Rabi flop time $T_\mrm{Rabi}=2\pi/(2g)$.
We obtain $T_\mrm{Rabi}=21.3~$ps and 5.36 ps for $\hbar g=100~\mu$eV and $400~\mu$eV, respectively.
Comparing these numbers we qualitatively understand why the cavity degrees of freedom can experience non-Markovian dephasing from the phonons for a large QD-cavity coupling, while the dephasing  essentially becomes Markovian for small QD-cavity coupling rates.
Alternatively, considering the spectral domain, one should compare the spectral features of the effective phonon spectrum, see \pref{fig:Dph_plots}, at the position and over the bandwidth of the polariton quasi-particle.
If the effective phonon spectrum is approximately constant in the vicinity of the polariton quasi-particle, the corresponding phonon timescale is much faster than the QD-cavity timescale and the cavity would interact in a Markovian fashion with the phonons.
On the other hand, if the phonon spectrum has significant variations across the polariton bandwidth, the two timescales will be comparable and the interaction will become non-Markovian.

While this discussion is not rigorous, it serves as a good rule-of-thumb for understanding and interpreting the results to be presented below.
We also note that present day state-of-the-art QD-cavity coupling constants \cite{Reinhard2011} are typically below $200~\mu$eV and thus in most experimentally relevant situations, we expect the cavity dephasing to be Markovian.
\subsection{Temperature dependence}
\begin{figure}[ht]
 \includegraphics[width=0.48\textwidth]{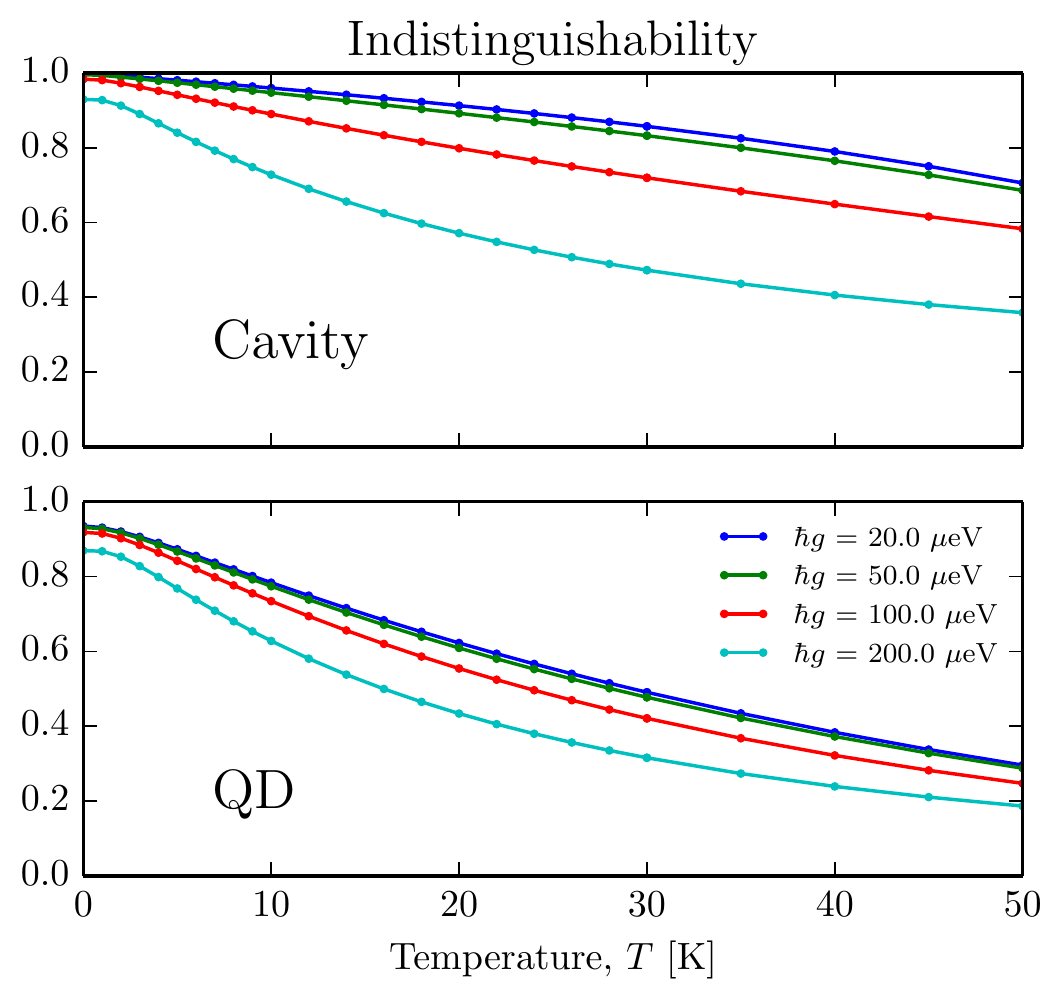}
 \caption{Temperature dependence of cavity (top) and QD (bottom) indistinguishability. Parameters: $\hbar \kappa=100~\mu$eV, $\hbar \Delta=0$, $\hbar \Gamma=1~\mu$eV, $\hbar \gamma=0$, $l_\mrm g = l_\mrm e=5~$nm.}
\label{fig:nmqrt_T_sweep}
\end{figure}

As seen from \pref{fig:Dph_plots}, the temperature is an important parameter and furthermore provides one of the few experimental handles for controlling the system after fabrication.
We have calculated the degree of indistinguishability as a function of temperature in the experimentally relevant range from $0$~K to $50$~K, see \pref{fig:nmqrt_T_sweep}.
We note that for elevated temperatures, phonons are thermally excited and the importance of multi-phonon processes is expected to increase, however our theory is limited to one-phonon processes due to its second order nature.
The results are therefore less accurate in the high temperature regime, but we note that previous studies \cite{Kaer2012c} show excellent agreement between a TCL second order theory and a multi-phonon theory up to at least $T=60$ K.

As expected, we observe the general trend of a monotonically decreasing degree of indistinguishability for both QD and cavity emission when the temperature is increased.
Comparing the results for QD and cavity emission, we observe, however, that the cavity indistinguishability depends more strongly on $g$ than is the case for QD indistinguishability.
As discussed in \pref{sec:puredephrates}, the phonon spectrum is sampled at $\om\approx 2g$ for the cavity case and higher phonon densities are thus obtained for larger $g$, leading to the smaller degree of indistinguishability.
For the QD case this is also true in the long-time limit, but the QD also suffers strong dephasing in the short-time regime, which tends to make the $g$-dependent contribution to dephasing less important.
\subsection{Phonon lifetime dependence}
As evidenced by \pref{eq:longtime_M22_coef}, the specific shape of the effective phonon density plays an important role in the decoherence induced by the phonons.
In the limit of small QD-cavity detuning and coupling, the behavior near $\om \approx 0$ of $d_\mathrm{ph}(\om)$, \pref{eq:eff_phonon_density_expli}, becomes important and is given by
\al{\label{eq:eff_phonon_density_expli_small_om}
d_\mathrm{ph}(\om)\approx\frac{(D_\mrm{e}-D_\mrm{g})^2}{4 \pi d c^5_\mrm{s}}k_\mrm B T \om^2.
}
Note that for very small frequencies, the frequency dependence of the effective phonon density is dominated by the nature of the deformation potential interaction, i.e., the square root factor $\sqrt{k}$ in \pref{eq:bare_phonon_ME}, the dimensionality of the problem, i.e., the integration volume element $k^2$, and the Bose function, i.e., $k^{-1}$ in the small frequency limit, whereas the form factor introduced by the finite sized QD wavefunctions plays no role.
The expression shows that $d_\mathrm{ph}(\om)$ approaches zero for $\om\rightarrow 0$, even at finite temperatures.
This has the consequence that in models where the deformation potential interaction is the only source of decoherence, near-unity indistinguishability is reached as $\sqrt{4g^2+\Delta^2}\rightarrow 0$ for cavity emission\cite{Kaer2013,Kaer2013a}.

\begin{figure}[ht]
 \includegraphics[width=0.48\textwidth]{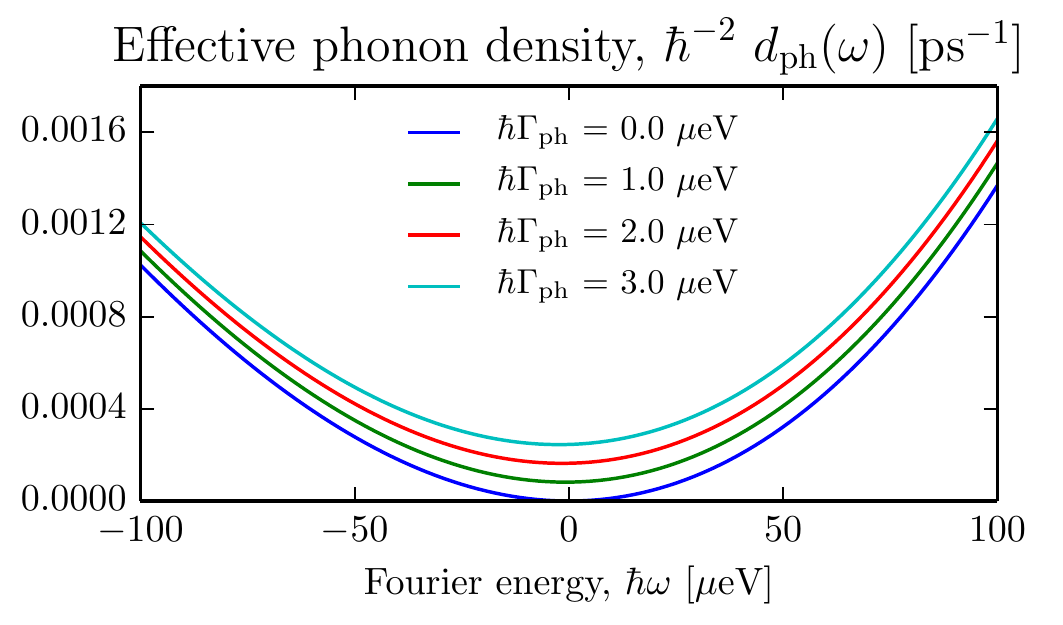}
 \caption{Effective phonon density for different values of the phonon lifetime phonon lifetime, \pref{eq:eff_phonon_density_gamLA}, focusing on the spectral region near $\om = 0$ where the relative effect of phonon decays is the largest is largest. Parameters: $T=4~$K and $l_\mrm g = l_\mrm e=5~$nm.}
\label{fig:twotime_puredephrate_gamLA_Dph}
\end{figure}

This property of the deformation potential interaction also implies a non-broadened zero phonon line (ZPL) in QD absorption spectra \cite{Krummheuer2002}, however it is well-known that several mechanisms lead to a finite width of the ZPL, e.g. the coupling to excited QD states \cite{Muljarov2004} and finite phonon lifetimes \cite{Stroscio2001,Zimmermann2002,Jacak2002,Hofmann2013} due to, e.g., surfaces, crystal impurities, or anharmonic interactions leading to decay of one phonon into two of smaller energy.
The simplest mechanism to implement in our model, and one that will always be present, is the finite lifetime of the phonons.
We include this as an overall exponential decay of the phonon correlation function \cite{Zimmermann2002} by a rate $\Gamma_\mrm{ph}$, i.e.
\al{\label{eq:Dph_def_gamLA}
D(t)=\sum_{\bs k} \abs{M^{k}}^2 \bS{n_{k}\e{+ i \omm{k}t - \Gamma_\mrm{ph}t}+\bP{n_{k}+1}\e{- i \omm{k}t- \Gamma_\mrm{ph}t}}.
}
Fourier transforming and taking the real part yields
\ml{\label{eq:eff_phonon_density_gamLA}
d_\mathrm{ph}(\om)=\sum_{\bs k}\abs{{M}_{k}}^{2}\\
\times[(n_{k}+1)L_{\Gamma_\mrm{ph}}(\om-\om_{k})+n_{k}L_{\Gamma_\mrm{ph}}(\om+\om_{k})],
}
which is nearly identical to \pref{eq:eff_phonon_density} except that the delta functions have been replaced with finite-width Lorentzians
\al{
L_\Gamma(\om)=\frac{\Gamma}{\om^2+\Gamma^2}.
}
The effect of a finite phonon lifetime is shown in \pref{fig:twotime_puredephrate_gamLA_Dph} in the vicinity of $\om\approx 0$, where the relative effect is the largest.
In contrast to \pref{fig:Dph_plots}, we now observe a finite phonon density at $\om = 0$, arising from the uncertainty in phonon energy induced by the finite lifetime.

\begin{figure}[ht]
 \includegraphics[width=0.48\textwidth]{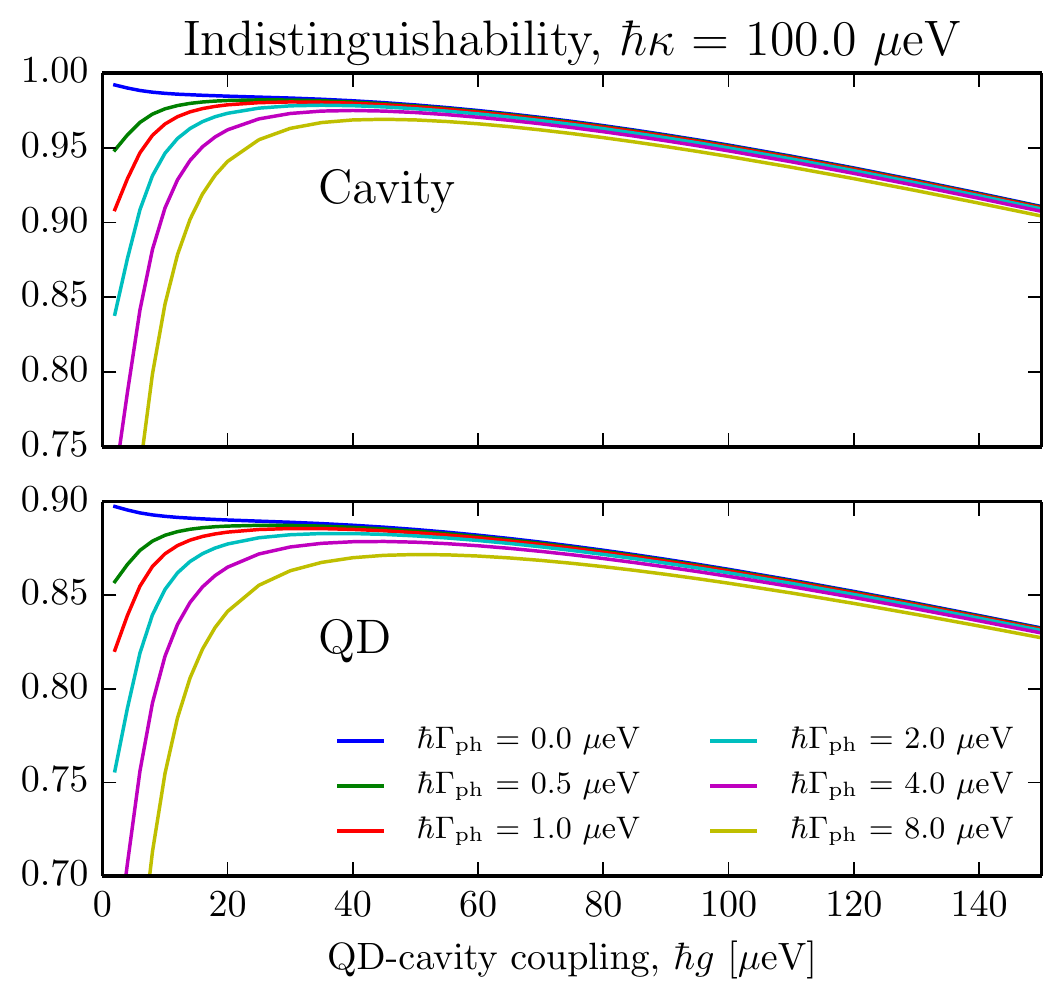}
 \caption{Dependence of cavity (top) and QD (bottom) indistinguishability on phonon lifetime as a function of QD-cavity coupling. Parameters: $\hbar \kappa=100~\mu$eV, $\hbar \Delta=0$, $\hbar \Gamma=1~\mu$eV, $\hbar \gamma=0$, $l_\mrm g = l_\mrm e=5~$nm, and $T=4~$K. See \pref{fig:NMQRT_Gam_ph_MORE_KAPPAS1} for more $\kappa$ values}
\label{fig:NMQRT_Gam_ph}
\end{figure}

To illustrate the effect of a finite phonon lifetime we show in \pref{fig:NMQRT_Gam_ph} the degree of indistinguishability as a function of the QD-cavity coupling $g$ for a range of phonon lifetimes \cite{Kaer2013a} ranging from 82.5 ps $(\hbar\Gamma_\mrm{ph}=8~\mu\mrm{eV})$ to $\infty$ for no decay $(\hbar\Gamma_\mrm{ph}=0)$.
The qualitative behavior is similar for the QD and cavity emission, with the absolute value for the QD indistinguishability always being significantly lower than for the cavity.
For $\Gamma_\mrm{ph}=0$ the curves monotonically decrease as a function of $g$ \cite{Kaer2013,Kaer2013a}, reflecting that even though the Purcell effect makes the photon emission faster (until we reach the strong coupling regime), the dephasing from the phonons also increases.
For finite phonon lifetimes the behavior changes qualitatively.
Instead of a monotonic decrease, we now observe an initial increase in indistinguishability for increasing $g$, followed by a maximum in the curve and a subsequent decrease.
The lower degree of indistinguishability for small $g$ arises due to the finite phonon density as $\om\rightarrow 0$, meaning that the long-time Markovian dephasing tends towards a finite value instead of zero as for $\hbar\Gamma_\mrm{ph}=0$.
Also, for small $g$, the effective QD decay rate is small, since the Purcell enhancement of the rate is small, and the photons have a longer timespan to experience dephasing.
The emergence of the maximum is a manifestation of the competition between the decoherence processes decreasing the indistinguishability and the Purcell effect which increases the indistinguishability by virtue of a faster QD decay.
We emphasize that the maximum in \pref{fig:NMQRT_Gam_ph} is not predicted by the standard Lindblad approach for including pure dephasing \cite{Kaer2013,Kaer2013a}, which lacks the mechanism increasing the phonon interaction for increasing $g$,  thus leading to the prediction of a ``saturation" of the indistinguishability for large $g$ rather than a maximum.
Considering that $\hbar\Gamma_\mrm{ph}=1~\mu$eV corresponds to a relatively long phonon lifetime of approximately $658~$ps, much longer than the typical correlation time of the phonon reservoir of $\sim$ 5 ps, and the relatively small quantitative effect it has on the overall effective phonon density, compare \pref{fig:Dph_plots} and \pref{fig:twotime_puredephrate_gamLA_Dph}, the large effect on the degree of indistinguishability is perhaps surprising.
However, the effect of the finite phonon lifetime correlates with the overall timescale of the QD-cavity system: The influence is seen to be largest for small $g$, where the system dynamics occurs on timescales comparable to the phonon lifetime and smaller for larger $g$, where the system dynamics is significantly faster than the phonon lifetime.

\begin{figure*}[ht]
 \includegraphics[width=0.48\textwidth]{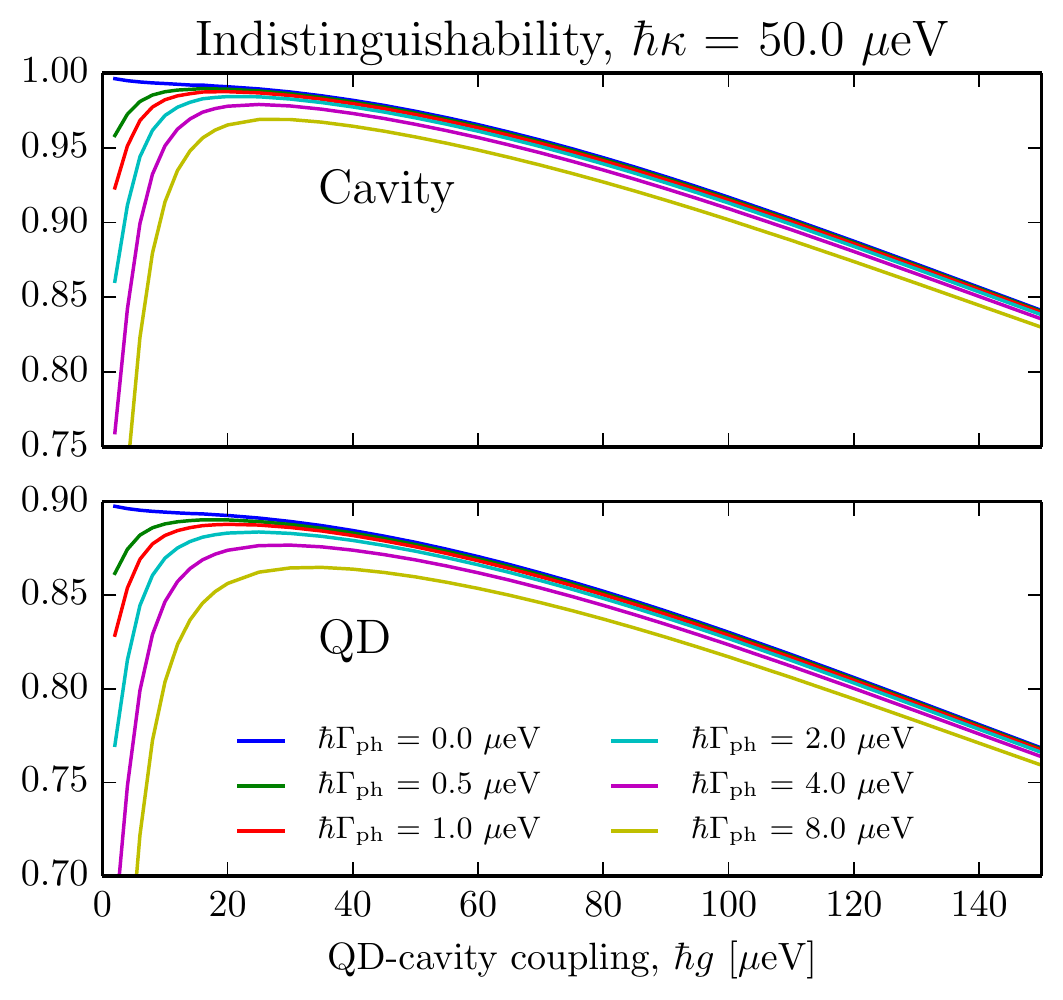}
 \includegraphics[width=0.48\textwidth]{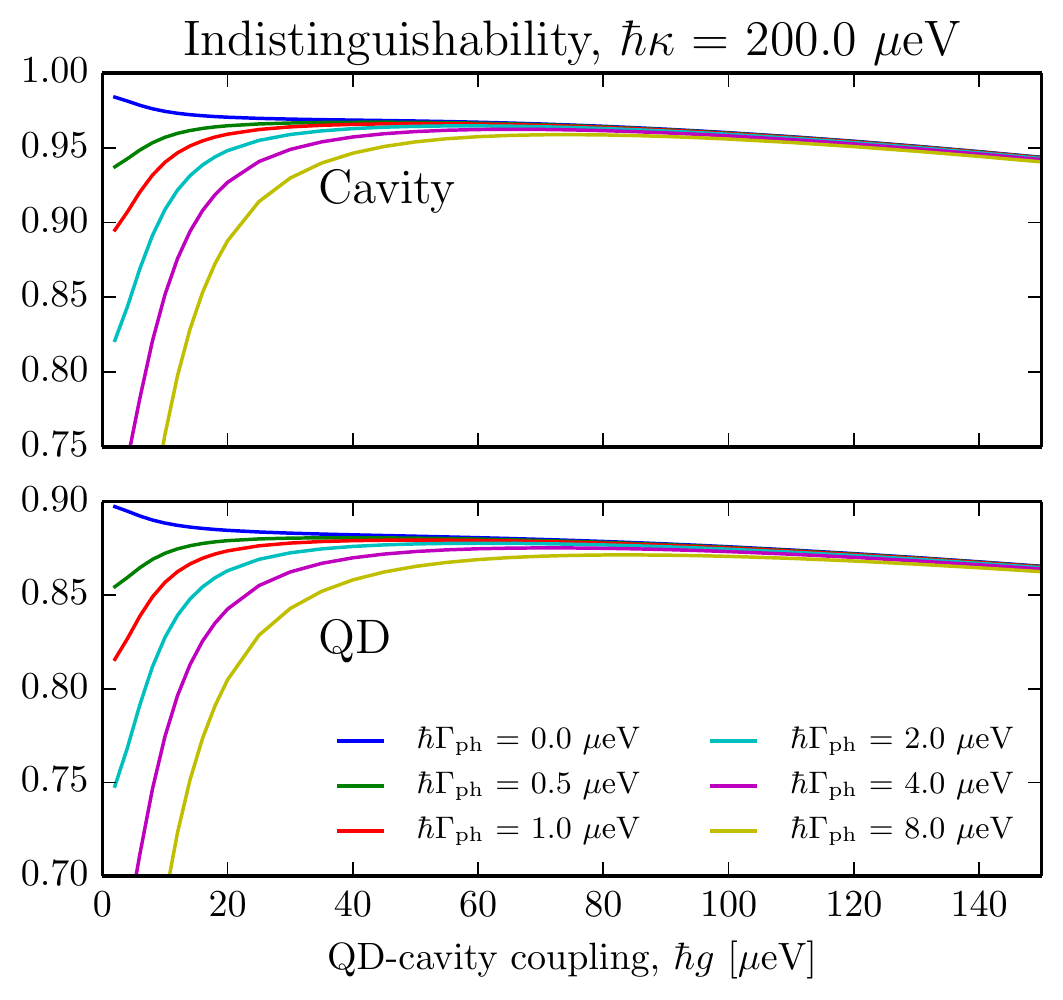}
 \caption{As \pref{fig:NMQRT_Gam_ph} but for additional $\kappa$ values.}
\label{fig:NMQRT_Gam_ph_MORE_KAPPAS1}
\end{figure*}
\subsection{Dependence on carrier confinement}
It is well-known that the QD wavefunctions, for both the ground and excited states, play an important role in the phonon interaction \cite{Krummheuer2002,Nysteen2013}.
This is illustrated in the lower panels of \pref{fig:Dph_plots} by the effective phonon density, where we varied the confinement lengths of the QD ground and excited state wavefunctions, $l_\mrm e$ and $l_\mrm g$, both independently and while keeping the two lengths identical.
The effective phonon density is seen to change when varying the confinement lengths over physically relevant values, and there is a general tendency for spatially confined (extended) wavefunctions to promote interaction with many (few) phonon modes, leading to a larger (smaller) density.
This can easily be inferred from the mathematical form of the phonon matrix element in \pref{eq:bare_phonon_ME}, which is proportional to the spatial Fourier transform of the QD wavefunction density.
\begin{figure}[ht]
 \includegraphics[width=0.48\textwidth]{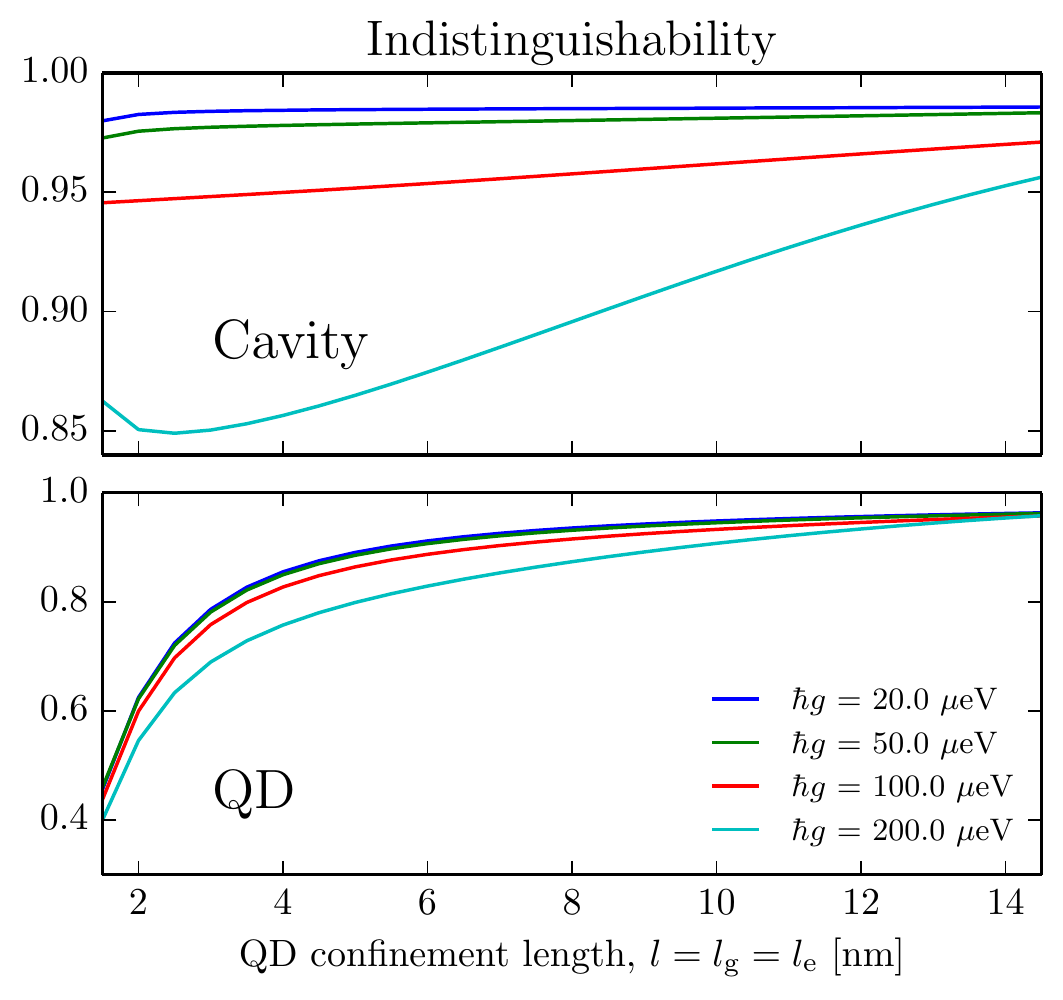}
 \caption{Variation of cavity (top) and QD (bottom) indistinguishability versus QD confinement lengths (assumed equal for excited and ground states, $l=l_\mrm e=l_\mrm g$). Parameters: $\hbar \kappa=100~\mu$eV, $\hbar \Delta=0$, $\hbar \Gamma=1~\mu$eV, $\hbar \gamma=0$, and $T=4~$K.}
\label{fig:WF_EQUALsize}
\end{figure}

In \pref{fig:WF_EQUALsize} we show the indistinguishability for QD and cavity emission as a function of confinement length $l=l_\mrm g =l_\mrm e$.

The indistinguishability for the QD emission is seen to depend strongly on the confinement length for all values of the QD-cavity coupling, and increases monotonically with the confinement length, with the strongest increase occurring for $l<5~$nm.
Comparing with the middle panel in \pref{fig:Dph_plots}, the behavior of the QD indistinguishability correlates well with the total number of available phonons, proportional to the integral over $d_\mrm{ph}(\om)$.
This dependence arises due to the dephasing acquired in the short-time non-Markovian regime, where the entire phonon bath is sampled.
The curve for $\hbar g=200~\mu$eV deviates slightly from the others, which is due to the significantly larger phonon density available at $\hbar\om\approx 2\hbar g=400~\mu$eV, which governs the long-time dephasing.

The indistinguishability for cavity emission shows a somewhat different behavior, with the dependence on $l$ being strongly dependent on $g$.
For small $g$, the indistinguishability only slightly increases with $l$, whereas for larger $g$ a much stronger increase is observed.
The weak dependence on $l$ for small QD-cavity coupling strength can be understood from \pref{eq:eff_phonon_density_expli_small_om}, which shows that for small $\om\approx 2 g$ the effective phonon density does not depend on the QD form factors and thus the confinements lengths.
Only for larger $\om\approx 2g$ does the confinement lengths start to affect the effective phonon density.

\begin{figure}[ht]
 \includegraphics[width=0.48\textwidth]{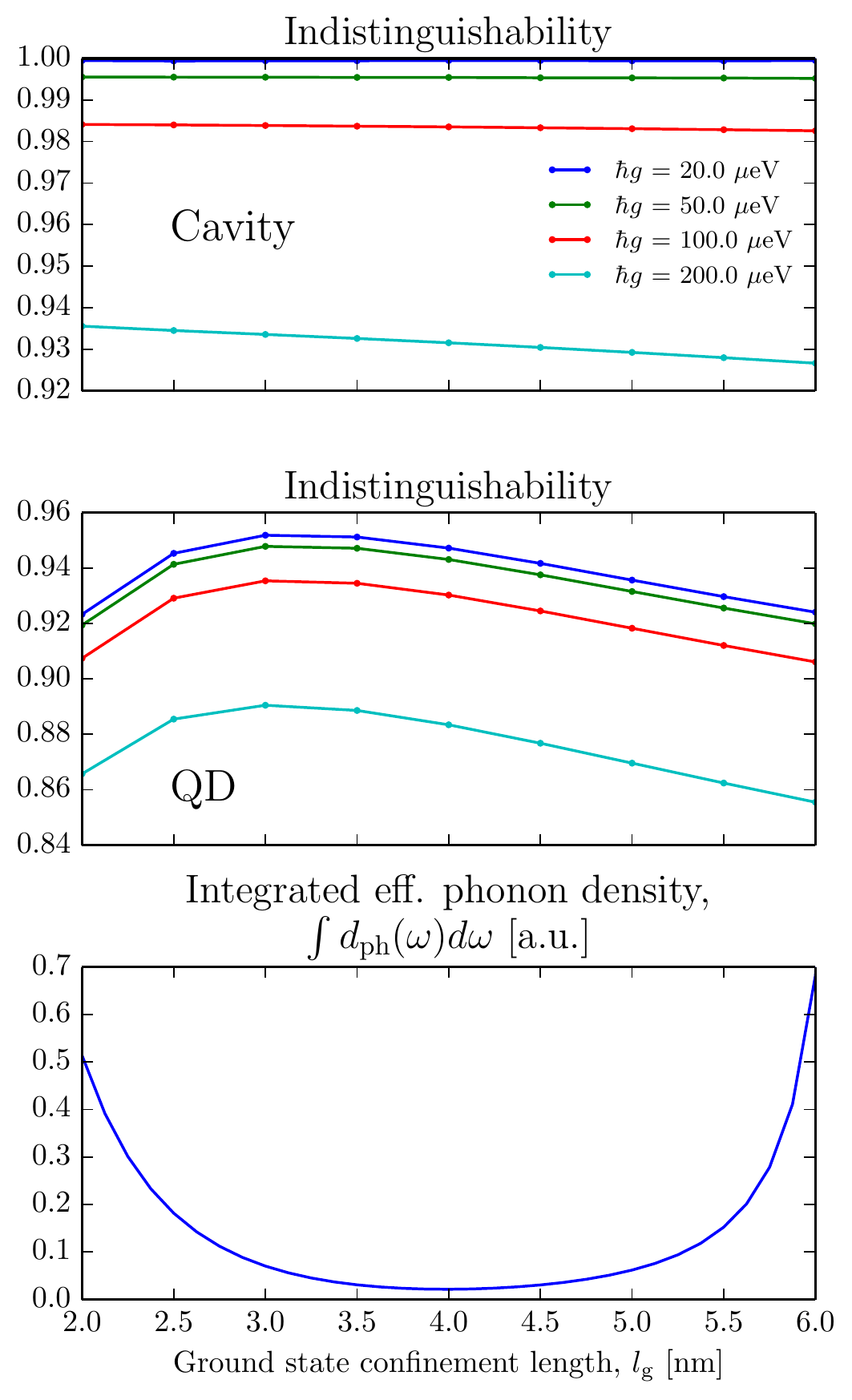}
 \caption{Dependence of cavity (top) and QD (middle) indistinguishability on confinement lengths. The excited state confinement length is chosen so that $l^3_\mrm e + l^3_\mrm g=2\times(5~\mrm{nm})^3$ is fixed. Frequency integral (bottom) over the  over the effective phonon density, \pref{eq:eff_phonon_density_expli}, $\int d_\mrm{ph}(\om)d\om$. Parameters: $\hbar \kappa=100~\mu$eV, $\hbar \Delta=0$, $\hbar \Gamma=1~\mu$eV, $\hbar \gamma=0$, and $T=4~$K.}
\label{fig:spectralhole_0detun}
\end{figure}

Recently, it was shown \cite{Nysteen2013} that the phonon interaction could be quenched by a suitable choice (or engineering) of the QD confinement lengths.
More specifically it was shown that when the QD wavefunctions for the ground and excited states were of unequal size, a compensation occurred leading to local minima in the effective phonon density at specific frequencies.
For certain classes of QDs the phonon density becomes zero at these frequencies, and phonon scattering is completely quenched.
This is illustrated in the bottom panel of \pref{fig:Dph_plots}, where we have kept the total volume of the ground and excited wavefunctions fixed, i.e. $l^3_\mrm g+l^3_\mrm e=\mrm{constant}$.
We see that the phonon density dips to zero in the interval between $1$ and $2$ meV for the different curves.
Importantly, the overall magnitude of the phonon density also decreases, meaning that the total number of phonons available for scattering has decreased, cf. the bottom panel of \pref{fig:spectralhole_0detun} showing the frequency integrated $d_\mrm{ph}(\om)$.
Furthermore, it was shown that for parameters such that $\sqrt{4g^2+\Delta^2}\sim\om_\mrm{dip}$ very small pure dephasing rates are obtained in the Markovian regime.
This result and the commonly used approximate expression \cite{Bylander2003,Kaer2013a} for the degree of indistinguishability, $I=\Gamma_\mrm{QD}/(\Gamma_\mrm{QD} + 2\gamma^*_\mrm{phonon})$, was used to speculate that one might improve the degree of indistinguishability by, e.g., adjusting the QD-cavity detuning to match the dips in the effective phonon density.
To investigate whether this suggestion holds true using a more accurate model, we calculate the effect of unequal QD confinement lengths on the degree of indistinguishability, cf. top panels of \pref{fig:spectralhole_0detun}.

For the cavity emission we observe the same trend as for $l_\mrm e=l_\mrm g$, see \pref{fig:WF_EQUALsize}, i.e. a weak dependence on confinement lengths for small $g$ and a stronger dependence for larger $g$.
Again the reason is that for small $\om\approx 2g$ the phonon density is weakly dependent on QD parameters, which only become important for larger $\om\approx 2g$.
In order to probe the dips in the effective phonon density, we would need $\sqrt{4g^2+\Delta^2}\approx 1-2~$meV, see \pref{eq:longtime_M22_coef}.
However, for state-of-the-art samples, $g$ remains relatively low and one would need $\hbar\Delta\approx 1-2~$meV to approach the spectral area where the density shows dips.
This in turns yields a very weak Purcell effect, leaving the system susceptible to dephasing and in the end the cavity indistinguishability does not benefit from the dips.

The above discussion has focused on the effect of varying the QD confinement lengths, however the spectral positions of the dips are not only determined by the confinement lengths, as the following expression shows \cite{Nysteen2013}
\al{
\om_\mrm{dip}^2=\frac{4c^2_\mrm s}{l^2_\mrm e-l^2_\mrm g}\mrm{ln}\bP{\frac{D_\mrm e}{D_\mrm g}}.
}
We note that especially the deformation potential constants display large experimental variations \cite{Vurgaftman2001}.

For QD emission we observe a much more interesting behavior, with a maximum occuring for intermediate values of $l_\mrm g$.
The confinement length for which the QD indistinguishability assumes a maximum, almost coincides with the confinement length for which a minimum is observed in the integrated phonon density.
This is not a coincidence, since we know that the QD indistinguishability is very sensitive to the overall magnitude of the effective phonon density, rather than values probed at specific frequencies, as for the cavity indistinguishability.
The shift of the maximum in QD indistinguishability towards smaller $l_\mrm g$, compared to the minimum of the integrated phonon density, is attributed to the smaller phonon density at small
$\om$, which is important for decoherence in the long-time limit, see bottom panel in \pref{fig:Dph_plots}.
We note that although we employ rather idealized spherically symmetric QD wavefunctions, spectral features similar to those discussed above are observed using QD models employing both more realistic confinement potential geometries and material compositions \cite{Nysteen2013}.
\subsection{Detuning dependence}
The QD-cavity detuning, $\Delta$, is an important parameter in cQED, as it is one of the few parameters that can be controlled externally during an experiment, typically by varying the temperature or via gas deposition on the photonic structure.
\begin{figure}[ht]
\includegraphics[width=0.48\textwidth]{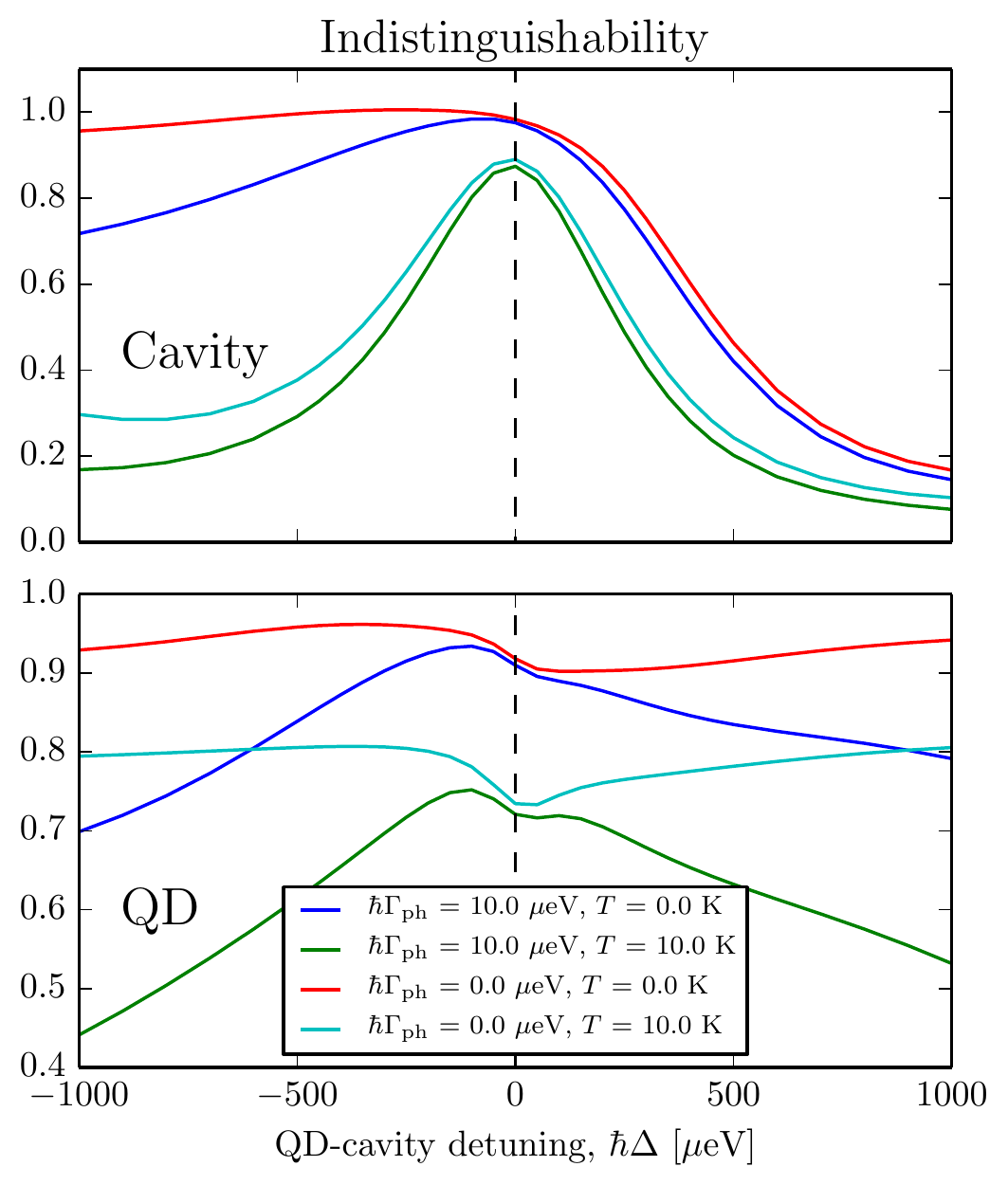}
 \caption{Dependence of cavity (top) and QD (bottom) indistinguishability on QD-cavity detuning $\Delta$. Parameters: $\hbar \kappa=100~\mu$eV, $\hbar g=100~\mu$eV, $\hbar \Gamma=1~\mu$eV, and $l_\mrm h = l_\mrm e=5~$nm.}
\label{fig:nmqrt_detun_Gam_ph}
\end{figure}

In \pref{fig:nmqrt_detun_Gam_ph} we show the degree of indistinguishability for QD and cavity emission as a function of QD-cavity detuning, $\Delta$, while varying the temperature and the phonon lifetime through $\Gamma_\mrm{ph}$.

Considering the cavity indistinguishability in the top panel of \pref{fig:nmqrt_detun_Gam_ph} for $\Gamma_\mrm{ph}=0$, we observe that the indistinguishability anti-correlates with the shape of the effective phonon density, \pref{fig:Dph_plots}.
For $T=0~$K the indistinguishability decreases strongly as the detuning approaches $1~$meV, where the phonon density has a maximum.
Further increasing the detuning, we expect the indistinguishability to recover, as it samples a decreasing density.
For negative detuning the indistinguishability approaches unity, mirroring the absence of thermally excited phonons, that would be responsible for a non-zero density for $\Delta<0$.
We note that near a detuning of $\hbar\Delta\sim -250~\mu$eV, the indistinguishability for cavity emission converges to a value of 1.0044, which is above unity and therefore unphysical.
This is a well-known problem associated with the TCL method, which in principle does not guarantee physical results but often works in practice, see discussions pp. 127-131 of Ref. \onlinecite{Breuer2002}.
Increasing the temperature to $T=10~$K causes an overall drop in the degree of indistinguishability, as expected, and now a significant reduction in indistinguishability is also observed for negative detuning, again mirroring the effective phonon density.
Including a finite phonon lifetime only slightly changes the results for $\Delta>0$, since here the relative change in phonon density is small.
However, for $T=0~$K the difference is significant, since now the phonon density is no longer strictly zero, but attains a small finite values due to the uncertainty in phonon energy induced by the phonon lifetime.
This causes the degree of indistinguishability to decrease for increasing absolute detuning, $|\Delta|$, as also observed in Lindblad models of pure dephasing \cite{Kaer2013}.
For $T=10~$K the influence of a finite phonon lifetime is only quantitative, but still significant,
consistent with the finite phonon density for $\Delta<0$.

In the bottom panel of \pref{fig:nmqrt_detun_Gam_ph} we show the indistinguishability for QD emission.
The curves for $T=0~$K and $10~$K display a very similar shape, except for an overall shift towards lower indistinguishability, in stark contrast to the cavity emission.
For $\Gamma_\mrm{ph}=0$, the individual curves tend to identical values as $|\Delta|\rightarrow\infty$, a behavior arising from the fact that the QD and cavity decouple for large detuning $|\Delta|$ and thus the QD indistinguishability will only depend on the phonon parameters and background QD decay, $\Gamma$.
However, for $\hbar\Delta< 1~$meV, the degree of indistinguishability still depends on the QD-cavity parameters and a smaller indistinguishability is observed for positive detuning, presumably due to the larger phonon density.
For $\Gamma_\mrm{ph}\neq 0$ the same trend as for the cavity emission is observed, i.e. the finite phonon lifetime induces a behavior similar to that observed for a constant pure dephasing rate, where the indistinguishability decreases as a consequence of the reduced Purcell effect.
%
%
%
\subsection{Useful approximations}\label{sec:useful_approximations}
The discussion in \pref{sec:puredephrates} indicates that certain simplifying approximations can be made to the full two-time EOM, \pref{eq:super_uber_twotime_EOM}, while still obtaining accurate results.
In this section we will investigate such approximations more systematically and quantify the error they introduce.
Starting by considering the scattering process most relevant for cavity emission, the approximate scattering term, \pref{eq:M22_cavity_approx}, can be obtained from the full expression, \pref{eq:M22_cavity}, by taking the Markovian long-time, i.e. $\tau\rightarrow\infty$ in the integration limits.
To get the remaining phonon induced scattering terms, not discussed in \pref{sec:puredephrates}, we take $\tau\rightarrow\infty$ in the integration limits of all terms in \pref{eq:super_uber_twotime_EOM}.
This leaves only terms from the first scattering integral on line two and three of \pref{eq:super_uber_twotime_EOM}, which are identical to those obtained in a treatment based on the Markovian quantum regression theorem \cite{Carmichael1999}.

For the scattering terms most relevant for QD emission, the approximate term, \pref{eq:M33_cavity_approx}, can be obtained from \pref{eq:M33_QD} by simply ignoring all terms in the last scattering integral on line four and five of \pref{eq:super_uber_twotime_EOM}.
Combining these observations it seems that both involve ignoring the last scattering integral in \pref{eq:super_uber_twotime_EOM}.
However, while in the cavity case the Markovian long-time limit is taken in the remaining scattering terms, the $\tau$ time dependence in the scattering terms is kept in the QD case.
The resulting scattering terms in the QD case are identical to those one would obtain, if one naively tried to take non-Markovian effects into account by using the non-Markovian one-time equations arising from the TCL, see e.g. \cite{Kaer2010}, and simply replaced the $t$-dependent scattering terms with $\tau$-dependent ones.

To summarize the above discussion, the approximative two-time EOM we are going solve is
\begin{widetext}
\ml{\label{eq:super_uber_twotime_EOM_approx}
\pd{\tau}\braket{A(t+\tau)B(t)} = \frac i\hbar\braket{\bT{[H_S,A]}(t+\tau)B(t)}\\
+\sum_{\nu_{1}\nu_{2}\nu_{3}\nu_{4}}\frac 1{\hbar^2}\int_{0}^{\tau_\mrm{max}}dt' \bS{ D^{*}_{\nu_{4}\nu_{3}\nu_{2}\nu_1}(t')  \braket{\{\ti P_{\nu_{1}\nu_{2}}(-t')[A,P_{\nu_{3}\nu_{4}}]\}(t+\tau)B(t)} \right.\\
+ \left. D_{\nu_{4}\nu_{3}\nu_{2}\nu_{1}}(t')\braket{\{[\hc P_{\nu_{3}\nu_{4}},A]\ti P^{\dagger}_{\nu_{1}\nu_{2}}(-t')\}(t+\tau)B(t)} },
}
\end{widetext}
where the upper integration limit takes different values.
For approximating the cavity terms we use $\tau_\mrm{max}=\infty$, which we will denote the Markov approximation, and for the QD case we use $\tau_\mrm{max}=\tau$, which we denote the naive approximation.

\begin{figure}[ht]
\includegraphics[width=0.48\textwidth]{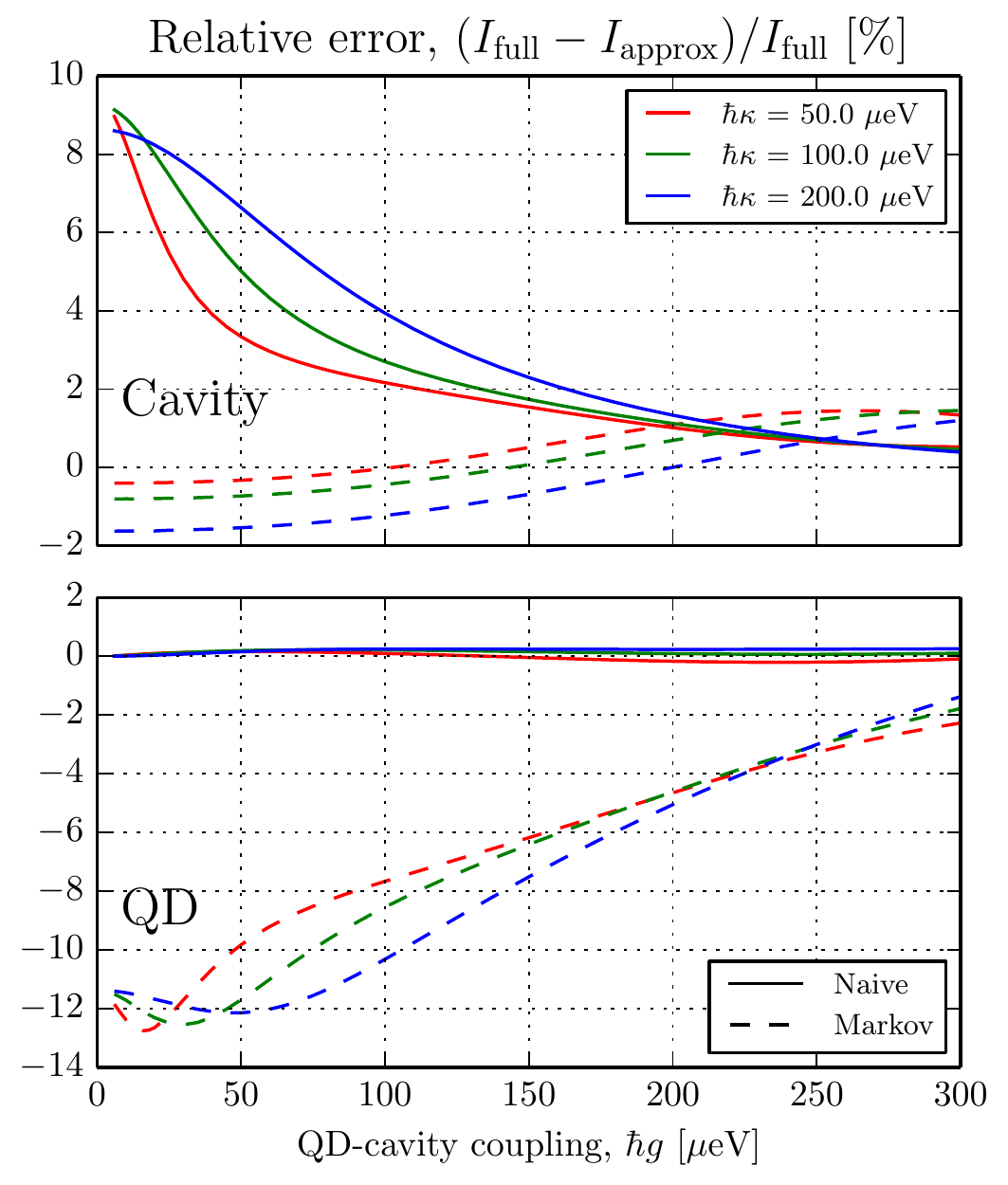}
 \caption{Relative error of the indistinguishability for various approximations, see main text. Parameters: $\hbar \Gamma=1~\mu$eV, $T=4~$K, $\hbar \Delta=0$, and $l_\mrm h = l_\mrm e=5~$nm.}
\label{fig:nmqrt_Approx_Compare_vs_g_and_kappa}
\end{figure}
In \pref{fig:nmqrt_Approx_Compare_vs_g_and_kappa} we compare the Markov and naive approximations with the full solution, by calculating the relative error as $\mrm{RE}=(I_\mrm{full}-I_\mrm{approx})/I_\mrm{full}$.
We investigate the dependence on the QD-cavity coupling $g$, which strongly changes the phonon-induced decoherence, for a series of experimentally relevant values of the cavity decay rate $\kappa$, as this parameter typically varies among different experiments.

For the case of the QD emission, the naive approximation is seen to give very good agreement with the full solution, and the relative error stays well below one percent for all considered parameters.
On the other hand, the Markov approximation leads to relative error larger than 10 \% for some parameters.

For cavity emission the situation is not as clear.
While the Markov approximation is clearly more accurate for most of the considered parameters, especially in the experimentally relevant region of relatively small $g$, the error depends strongly on the value of $\kappa$, and the relative error quickly becomes larger than one percent.

From a computational point of view, the Markov approximation is highly advantageous, since the resulting system of equations becomes time-independent and can be solved very efficiently using standard methods from linear algebra.
The naive approximation is not nearly as convenient as the Markov approximation, as the phonon-induced scattering rates retain their time-dependence, although only in $\tau$, and thus general time-stepping schemes must be employed to solve for the dynamics.
The structure of the problem does, however, allows for parallelization.  For each value of $t$, the corresponding EOM in $\tau$ thus decouple completely from the rest.
This is also the case for the full set of equations.
\subsection{Emission spectra}
The optical emission spectra provide important information about the cQED dynamics and have been studied extensively, both theoretically and experimentally.
Emission spectra were thus used to provide the first experimental demonstrations of strong coupling in a semiconductor cQED system \cite{Reithmaier2004,Yoshie2004}.
As for phonon effects on emission spectra, the inherent asymmetric spectral properties of phonons at low temperatures is to some degree transferred to the emission spectra, where incoherently \cite{Hohenester2010,Calic2011,Hughes2011} and coherent \cite{Weiler2012,Roy2011b,McCutcheon2013} pumped system have been investigated.
To calculate the emission spectra, here defined as \cite{Carmichael1999},
\al{\label{eq:EMSPEC}
S(\om_\mrm S) \propto \mrm{Re}\bS{ \int_{-\infty}^{+\infty}dt \int_{0}^{+\infty}d\tau \braket{\hc A(t+\tau)A(t)}e^{-i\om_\mrm S \tau}}
}
the first order two-time correlation function $\braket{\hc A(t+\tau)A(t)}$ is needed, which also enters into the definition of the degree of indistinguishability, \pref{eq:ID_def}.
Common to the studies mentioned above, the QRT has been used to calculate the two-time correlation function, thus implying that the Markov approximation has been enforced.

It was demonstrated earlier \cite{Kaer2013,Kaer2013a}, and discussed at length in the present work, that non-Markovian effects due to the interaction with phonons can give rise to large deviations compared to a Markovian treatment when calculating the degree of indistinguishability.
Since both the emission spectra and the degree of indistinguishability depend on the two-time correlation function, one might worry about the validity of employing a Markovian framework to determine the emission spectra in the presence of phonon interactions. We are not aware of any studies comparing emission spectra calculated within the QRT to the emission spectra obtained when non-Markovian effects are taken into account.
\begin{figure}[ht]
 \includegraphics[width=0.48\textwidth]{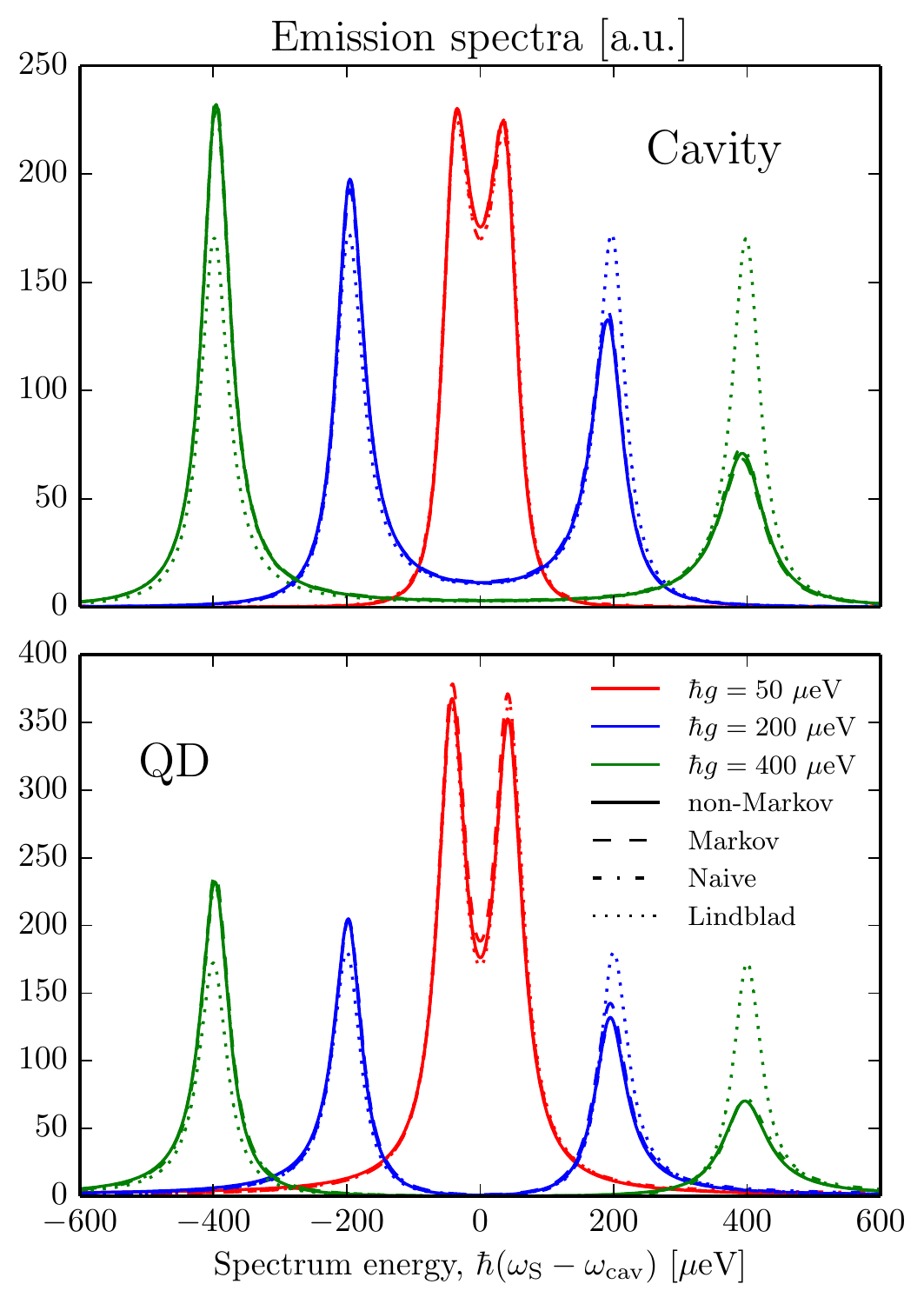}
 \caption{Emission spectra, \pref{eq:EMSPEC}, comparing the NM2PT with various approximations. Parameters: $\hbar \Delta=0$, $\hbar \Gamma=1~\mu$eV, $\hbar \kappa=100~\mu$eV, $\hbar \gamma=0$, $l_\mrm g = l_\mrm e=5~$nm, and $T=4~$K.}
\label{fig:EM_spec}
\end{figure}

To investigate this further we show in \pref{fig:EM_spec} the emission spectra for light emitted from QD and cavity, calculated using the NM2PT, the Markov and naive approximations discussed in the previous section, and finally a pure Lindblad model where phonon effects are not included, not even as a pure dephasing rate.
We have chosen parameters placing the system in the strong coupling regime, where the eigenstates are the upper (u) and lower (l) polariton branches with energies $\om_\mrm{u/l}\sim \om_\mrm{cav} \pm g$.
The first thing we notice is the asymmetry, with more light being emitted at frequencies below the common resonance frequency of the QD and the cavity, than at frequencies above it.
The asymmetry arises since the QD initially is in the excited state with the cavity being in the ground state. This corresponds to the state of the QD-cavity system being in an equal mix of the upper and lower polariton branches, $\sim \ket{\mrm u}+\ket{\mrm l}$.
In principle the system can make transitions both up and down in energy between the two branches by phonon emission and absorption, but since we are in a low-temperature regime, phonon emission processes are more likely to occur than absorption processes \cite{Hughes2011}.
This means that we are more likely to find the system in the low energy state, i.e. the lower polariton branch at $\om_\mrm{l}\sim \om_\mrm{cav} - g$, which consequently emits more light.
The degree of asymmetry is observed to increase as the QD-cavity coupling strength becomes larger, which is a consequence of more phonon modes being available at the higher phonon energies required to bridge the gap between the upper and lower branches of the polariton, see \pref{fig:Dph_plots}.
We note that the curves employing approximate Lindblad terms remain completely symmetric.

When comparing the curves obtained using the full non-Markovian theory and those using the Markov and naive approximations, we find only relatively minor deviations over the entire bandwidth of the spectra.
As discussed in Sections \ref{sec:puredephrates} and \ref{sec:useful_approximations}, the Markov approximation works well in predicting the indistinguishability for light emitted from the cavity and the so-called naive approximation works well for the indistinguishability of light emitted from the QD. On the other hand, the Markov approximation worked very poorly for QD emission and the naive approximation quite poorly for cavity emission.
It thus appears surprising that the approximations work so well for the emission spectra, while they do not work very well for the indistinguishability.
One reason may be that the emission spectra do not probe energies that are large enough to be affected by the short-time non-Markovian dynamics, which roughly corresponds to the peak in the effective phonon density, i.e. in the $1$ meV range.
\section{Summary and conclusions}\label{sec:summary}
In this paper we have employed a novel theory \cite{Goan2010,Goan2011a} to investigate the degree of indistinguishability of single photons emitted from semiconductor cQED systems when subject to dephasing due to scattering with longitudinal acoustical (LA) phonons.
In particular, we have accounted for the non-Markovian aspects of the interaction with the phonon reservoir and performed extensive investigations of how the photon indistinguishability depends on the parameters of the system.
The following is a summary of the obtained results and conclusions.

In \pref{sec:theory} we started by introducing the perturbational second order theory (NM2PT) used for calculating the two-time functions for observables belonging to the cQED system.
The derivation is based on the timeconvolution-less formalism and thus the resulting equations do not contain explicit memory integrals, but rather encode the non-Markovian evolution in time-dependent scattering rates.
In the framework of this theory we defined the Markovian and non-Markovian regimes in terms of the relevant timescales.

Then we presented the mathematical models describing the cQED system consisting of the QD and cavity, namely the Jaynes-Cummings model, and the bulk model representing the continuum of LA phonon modes, coupled to the cQED system via the deformation potential interaction.
Furthermore we introduced the effective phonon density, governing the phonon modes available for scattering with the QD, while taking into account the effect of finite temperatures.
This quantity plays an extremely important role in interpreting the effect of phonons on the degree of indistinguishability and is extensively employed in all parts of the paper.

In \pref{sec:results} we started out by considering the parameter regime of large cavity losses, typically where $\kappa > g$, which is commonly encountered in experiments.
To estimate the accuracy of the NM2PT in this regime, we compared to an exact diagonalization approach \cite{Kaer2013,Kaer2013a} without any uncontrolled approximations.
We found that the perturbation theory breaks down for $\hbar\kappa > 300~\mu$eV, for typical experimentally relevant parameters.
We also derived a simple analytical expression for the indistinguishability in the large $\kappa$ limit.
The break-down of the NM2PT is speculated to arise from the absence of non-perturbative effects of a large cavity loss in the phonon scattering rates, which only include coherent properties of the system, i.e. the Jaynes-Cummings system is assumed lossless.
This is a consequence of the basic assumptions of the theory, which become important whenever the loss rates become larger than the coherent system parameters.
An \emph{ad hoc} fix was investigated, involving manually including loss in the scattering terms, however no systematic improvements of the results was found.

A discussion of the two-time scattering rates arising from the coupling to the non-Markovian reservoir was presented, with the goal of establishing when the system-reservoir interaction is of Markovian or non-Markovian type.
It was found that the importance of a full non-Markovian treatment depends on whether one considers light emitted from the cavity or from the QD.
In the regime of small QD-cavity coupling strength, the degree of indistinguishability of light from the QD requires a non-Markovian treatment, while this was not the case for cavity light, which could be described in the Markovian limit.
In the regime of large QD-cavity coupling strength, both QD and cavity required a non-Markovian treatment.
The reason for this behavior is the inability of the QD-cavity coupling to mediate the fast phonon dynamics, on the order of the reservoir correlation time $\sim 5$ ps, from the QD to the cavity, unless the QD-cavity coupling time (inverse rate) is comparable to the phonon reservoir correlation time.

The dependence on temperature was also investigated and it was found that increasing the temperature causes a decrease in indistinguishability, for both QD and cavity emission.
This is due to the increased population of thermally activated phonons, which enables both absorption and stimulated emission processes that increase the scattering phase space and hence increase the rate of decoherence.
Emission from the QD was found to be more sensitive to temperature compared to the cavity, as the QD typically samples a much larger part of the available phonon modes than the cavity.

Models only including dephasing by phonon scattering predict a near-unity indistinguishability for the cavity emission in the regime of small QD-cavity coupling strengths.
This is due to the absence of available phonon modes for scattering near zero phonon frequency.
However, if the temporal decay of phonons is taken into account, the uncertainty in lifetime translates into an uncertainty in energy and a finite phonon density is sampled at zero phonon energy.
This was found to strongly affect the indistinguishability for both QD and cavity in the small $g$ limit, which is very relevant for experiments, and an optimum value of the QD-cavity coupling was predicted.

We showed how a strong carrier confinement gave rise to an, in general, smaller degree of indistinguishability, whereas a weak carrier confinement increased the indistinguishability.
Again, the QD emission was found to be more sensitive towards the degree of carrier confinement, due to its more non-Markovian behavior.
We also investigated the effect of varying the carrier confinement for excited and ground states independently, while keeping the total volume fixed, which has been shown to induce dips in the effective phonon density.
While the cavity emission was relatively insensitive towards this, the indistinguishability of QD emission was found to correlate with the overall number of available phonon states and a maximum degree of indistinguishability was predicted.

We investigated the dependence on the detuning and found that for the cavity emission, the indistinguishability correlated with the effective phonon density, and at elevated temperatures the effect of phonon absorption was very clear.
The degree of indistinguishability for QD emission was much less sensitive to detuning, since for large detuning the QD and cavity decouple and while energy must be transferred from the QD to the cavity for the cavity to emit light, the QD does not require the cavity in order to emit light.
\begin{acknowledgments}
This work was funded by Villum Fonden via the Centre of Excellence ``NATEC" and by the European Metrology Research Programme (EMRP) via the project SIQUTE (contract EXL02).
\end{acknowledgments}

\appendix
\section{Equation of motion for reduced density matrix}\label{app:EOM_RDM}
In this appendix we present the full set of dynamical equations used in the main text.
The time-evolution operator governing the interaction picture, \pref{eq:int_picture_def}, with respect to the system Hamiltonian, \pref{eq:Hs_def_first}, is explicitly given by
\al{
\label{eq:U_op_explicit}
U_\mrm S(t)&=e^{-iH_\mrm S t/\hbar}\notag\\
&=e^{-\frac 12 i \Delta t}\bmm{
\cos (\frac{\Omega t}{2})-\frac{i\Delta \sin (\frac{\Omega t}{2})}{\Omega}&-\frac{2ig \sin (\frac{\Omega t}{2})}{\Omega}&0\\
-\frac{2ig \sin (\frac{\Omega t}{2})}{\Omega}&\cos (\frac{\Omega t}{2})+\frac{i\Delta \sin (\frac{\Omega t}{2})}{\Omega}&0\\
0&0&1\\
},
}
where $\Omega=\sqrt{4g^2+\Delta^2}$.

The dynamical equations are derived from \pref{eq:super_uber_twotime_EOM}.
In the one-time case we take $B=I$ and $t=0$, for which the general two-time equation reduces to the correct one-time equation.

It is convenient to represent the one-time functions in a vector form as
\al{
\braket{\bs u(t)}=
\bmm{
\asig{11}\\
\asig{22}\\
\asig{12}\\
\asig{21}\\
},
}
in which case the set of equations can be written as
\al{\label{eq:onetime_EOMs}
\pd t\braket{\bs u(t)} = M(t)\braket{\bs u(t)},
}
where the time-dependent coupling matrix $M(t)$ is given by
\begin{widetext}
\al{
M_{1,1}(t)&=-\Gamma,\\
M_{1,2}(t)&=0,\\
M_{1,3}(t)&=-ig,\\
M_{1,4}(t)&=ig,\\
M_{2,1}(t)&=0,\\
M_{2,2}(t)&=-\kappa,\\
M_{2,3}(t)&=ig,\\
M_{2,4}(t)&=-ig,\\
M_{3,1}(t)&=-ig+H_{11,12}(t,0),\\
M_{3,2}(t)&=ig-H^*_{12,11}(t,0),\\
M_{3,3}(t)&=i(\Delta-\Delta_\mrm{pol})-\frac 12(\Gamma+\kappa+2\gamma)+\bT{H_{12,12}(t,0)-H^*_{11,11}(t,0)},\\
M_{3,4}(t)&=0,\\
M_{4,1}(t)&=ig+H^*_{11,12}(t,0),\\
M_{4,2}(t)&=-ig-H_{12,11}(t,0),\\
M_{4,3}(t)&=0,\\
M_{4,4}(t)&=-i(\Delta-\Delta_\mrm{pol})-\frac 12(\Gamma+\kappa+2\gamma)+\bT{H^*_{12,12}(t,0)-H_{11,11}(t,0)}.
}
\end{widetext}
Here we have subtracted the phonon-induced energy shift $\Delta_\mrm{pol}=-[H_{12,12}(\infty,0)-H^*_{11,11}(\infty,0)]=\hbar^{-2}\mrm{Im}\bS{\int_0^\infty dt D(t)}$ (often referred to as the polaron shift) from the QD-cavity detuning.
This is done in order to ensure that zero detuning, $\Delta=0$, corresponds to the cavity and QD being resonant.
We note that these equations are identical to those obtained in the conventional TCL for the same Hamiltonian, see e.g. Ref. \onlinecite{Kaer2010}.

As for the one-time functions, we represent the two-time correlation functions as a vector
\al{
\braket{\bs v(t+\tau,\tau)}=
\bmm{
\ttfct{\sigg{23}}{\sigg{32}}\\
\ttfct{\sigg{13}}{\sigg{32}}\\
\ttfct{\sigg{13}}{\sigg{31}}\\
\ttfct{\sigg{23}}{\sigg{31}}\\
}.
}
The EOMs can then be written as
\al{\label{eq:twotime_EOMs}
\pd\tau\braket{\bs v(t+\tau,\tau)} = M(t,\tau)\braket{\bs v(t+\tau,\tau)}
}
where the two-time dependent coupling matrix is given by
\begin{widetext}
\al{
M_{1,1}(t,\tau)&=-\kappa/2,\\
M_{1,2}(t,\tau)&=ig,\\
M_{1,3}(t,\tau)&=0,\\
M_{1,4}(t,\tau)&=0,\\
M_{2,1}(t,\tau)&=ig-\bT{H^*_{12,11}(\tau,0)+H^*_{12,11}(t+\tau,\tau)},\\
\label{eq:twotime_M22}
M_{2,2}(t,\tau)&=i(\Delta-\Delta_\mrm{pol})-\frac 12(\Gamma+2\gamma)-\bT{H^*_{11,11}(\tau,0)+H^*_{11,11}(t+\tau,\tau)}+G_{12,12}(t+\tau,\tau),\\
M_{2,3}(t,\tau)&=G_{11,12}(t+\tau,\tau),\\
M_{2,4}(t,\tau)&=0,\\
M_{3,1}(t,\tau)&=0,\\
M_{3,2}(t,\tau)&=G_{12,11}(t+\tau,\tau),\\
\label{eq:twotime_M33}
M_{3,3}(t,\tau)&=i(\Delta-\Delta_\mrm{pol})-\frac 12(\Gamma+2\gamma)-\bT{H^*_{11,11}(\tau,0)+H^*_{11,11}(t+\tau,\tau)}+G_{11,11}(t+\tau,\tau),\\
M_{3,4}(t,\tau)&=ig-\bT{H^*_{12,11}(\tau,0)+H^*_{12,11}(t+\tau,\tau)},\\
M_{4,1}(t,\tau)&=0,\\
M_{4,2}(t,\tau)&=0,\\
M_{4,3}(t,\tau)&=ig,\\
M_{4,4}(t,\tau)&=-\kappa/2.
}
\end{widetext}
The phonon-induced scattering rates are defined as
\al{\label{eq:H_coef_def}
H_{kl,nm}(t_1,t_2)&=\hbar^{-2}\int^{t_1}_{t_2}dt'W_{kl,nm}(t')D(t')\\
G_{kl,nm}(t_1,t_2)&=\hbar^{-2}\int^{t_1}_{t_2}dt'W_{kl,nm}(t'-t_2)D(t')\\
W_{nm,kl}(t)&=U^*_{nm}(t)U_{kl}(t).
}

%


\bibliographystyle{plain}

\bibliography{NMQRT_paper.bib,footnotes.bib}


\end{document}